\documentclass[a4paper,11pt]{article}
\pdfoutput=1 % if your are submitting a pdflatex (i.e. if you have
\usepackage{jheppub}
\usepackage[T1]{fontenc}
\usepackage{cancel}
\usepackage{subfigure}
\usepackage{xcolor}
\usepackage{slashed}
\usepackage{bbold}
\usepackage{graphics}
\usepackage{ dsfont }

\newcommand{\eq}[1]{\begin{align}#1\end{align}}

\renewcommand{\u}[1]{\textrm{U}(#1)}
\newcommand{\su}[1]{\textrm{SU}(#1)}
\newcommand{\so}[1]{\textrm{SO}(#1)}
\newcommand\T{\rule{0pt}{2.6ex}}       % Top strut
\newcommand\B{\rule[-1.2ex]{0pt}{0pt}} % Bottom strut
\newcommand{\nn}{\nonumber} 

\newcommand{\ba}{\begin{array}}
\newcommand{\ea}{\end{array}}
\newcommand{\disp}{\displaystyle}
\newcommand{\Lag}{{\cal L}}
\newcommand{\orden}{\sim}
\newcommand{\dis}{\hspace{0.08cm}}
\newcommand{\vev}{{\em vev}}
\newcommand{\hc}{{\rm h.c.}}

\title{Phenomenological implications of the new Littlest Higgs model with T-parity}

\author{Jos\'e Ignacio Illana and Jos\'e Mar\'ia P\'erez-Poyatos}

\affiliation{CAFPE and Departamento de F\'isica Te\'orica y del Cosmos, 
             Universidad de Granada, \\ E-18071 Granada, Spain}

\emailAdd{jillana@ugr.es, jmppoyatos@ugr.es}

\abstract{
We investigate the parameter space of the new Littlest Higgs model with T-parity (NLHT) recently introduced to cure some pathologies of the original LHT. The model requires extra fermion content and additional pseudo-Goldstone bosons. While the heavy top quark sector is similar, there are both T-odd and T-even heavy quarks and leptons with masses proportional to just two sets of Yukawa matrices in flavor space, one more than in the LHT. The new scalars are a singlet and real triplet, T-odd, with masses controlled by gauge and Yukawa couplings, independent of the spontaneous symmetry breaking scale $f$, and hence potentially light. Imposing that no mass exceeds the cutoff scale, applying current lower bounds on vector-like quarks and assuming a simplified model with mass degenerate heavy fermions compatible with the heavy photon as dark matter constituent, we find that $f$ gets constrained within the interval between 2 and 3~TeV, the common Yukawa coupling of heavy leptons gets fixed and the Yukawa coupling of heavy quarks becomes greatly correlated to the top quark Yukawa couplings. The particle spectrum is then bounded from below and above, with the (lightest) heavy photon at about 0.5~TeV, not far from the heavy leptons, the new scalars below 1~TeV, the usual complex scalar triplet close to the heavy weak bosons at about 1.5 to 2.5~TeV, and the heavy quarks and top quark partners between 2 and 5~TeV. The new scalars decay predominantly to a standard and a T-odd lepton and have a width comparable to that of the Higgs.
}
\begin{document}
%\texttt{[{\footnotesize Last version: \today}]}
\maketitle

\flushbottom

%-----------------------------------------------------------------------%
%--------------- 	INTRODUCTION -----------------------------------------%
%-----------------------------------------------------------------------%

\section{Introduction}\label{Introduction}
Composite Higgs models \cite{panicoCompositeNambuGoldstoneHiggs2015} are well motivated frameworks to study physics beyond the Standard Model (SM). In this family of models, the Higgs boson and, typically, also additional scalar degrees of freedom arise as pseudo-Nambu-Goldstone bosons (pNGBs) of a spontaneously broken approximate global symmetry \cite{colemanStructurePhenomenologicalLagrangians1969,callanStructurePhenomenologicalLagrangians1969}. This global symmetry gets explicitly broken by gauge and Yukawa interactions, so the pNGBs acquire a mass proportional to the symmetry breaking couplings at one loop through the Coleman-Weinberg potential \cite{colemanRadiativeCorrectionsOrigin1973}.

Within this class of models, the Littlest Higgs model with T-parity (LHT) \cite{arkani-hamedConstructingDimensions2001,arkani-hamedElectroweakSymmetryBreaking2001,ArkaniHamed:2002qy,chengTeVSymmetryLittle2003,chengLittleHierarchyLittle2004,lowParityLittlestHiggs2004,chengTopPartnersLittle2006} is one of the most elegant proposals. It is based on the global group \su5 broken spontaneously to \so5 by the vacuum expectation value ({\it{vev}}) of a symmetric tensor at the scale $f\orden 1-10$ TeV. The spontaneous breaking of the global group generates 14 pNGBs. The discrete T-parity symmetry is implemented to relax constraints from electroweak precision data (EWPD) \cite{hubiszPhenomenologyLittlestHiggs2005,hubiszElectroweakPrecisionConstraints2006}. To that end, all the SM particles are T-even and most of the new particles are T-odd and hence pair-produced. A subgroup $\left[\su2\times \u1\right]_1\times \left[\su2\times \u1\right]_2$ of the full \su5 is gauged. This gauge group gets spontaneously broken to the diagonal $\left[\su2\times \u1\right]$ giving rise to a set of massive T-odd gauge bosons $W_H^{\pm}$, $Z_H$ and $A_H$. In particular, the $A_H$ gauge boson is the lightest T-odd particle of the LHT spectrum and is electrically neutral, constituting a dark matter candidate \cite{hubiszPhenomenologyLittlestHiggs2005,2006,Wang:2013yba,Wu:2016rwz}. On the other hand, the Goldstone sector includes the T-even physical Higgs and the would-be Goldstone bosons absorbed by the SM gauge bosons, together with extra T-odd scalars: the would-be Goldstone bosons eaten by the heavy gauge bosons and a physical complex triplet. The Little Higgs (with and without T-parity) phenomenology has been extensively explored \cite{hubiszPhenomenologyLittlestHiggs2005,hubiszElectroweakPrecisionConstraints2006,2006,Dey:2008dk,Yue:2008zp,Wang:2013yba,Reuter:2013iya,Yang:2014gca,Aranda:2015rya,Kirca:2015iaa,Cao:2015cdb,Choudhury:2016wwt,Aranda:2017raf,Aranda:2017bgq,Wang:2017urv,Yang:2018oek,Yang:2018obf,Dercks:2018hgz,Long:2019oev,Yang:2019uea}, also providing new sources of quark and lepton flavor violating processes \cite{Belyaev:2006jh,Blanke:2006sb,Tarantino:2006afb,Blanke:2006eb,Duling:2007sf,Tarantino:2007ur,delAguila:2008zu,delAguila:2010nv,Blanke:2015wba,Buras:2015hna,delaguilaLeptonFlavorChanging2017,delaguilaFullLeptonFlavor2019,Pacheco:2021djh,Pacheco:2022ebc} and neutrino mass generation \cite{delAguila:2005yi,Han:2005nk,Lee:2005kd,Blanke:2007db,Hektor:2007uu,deAlmeida:2007khx,delaguilaInverseSeesawNeutrino2019}.

However, the LHT has been shown to suffer from pathologies in the fermionic sector. The fermion content breaks explicitly the gauge invariance of the LHT \cite{pappadopuloTparityItsProblems2011, illana2022new}. In this model, the left-handed components of the SM and the {\it{mirror}} fermions come in incomplete \su5 multiplets  whereas their right-handed counterparts come in a multiplet of \so5 \cite{Belyaev:2006jh,Blanke:2006sb,Blanke:2006eb,delAguila:2008zu,delAguila:2010nv,gotoTauMuonLepton2011,zhouFlavorChangingTop2013,yangLeptonFlavorViolating2017}. The latter contains an \su2 doublet of mirror fermions, together with an \su2 doublet of {\it{mirror partners}} and a gauge singlet
\cite{chengLittleHierarchyLittle2004,delaguilaLeptonFlavorChanging2017}. The mirror fermions acquire a vector-like mass of order $f$ by a Yukawa coupling to the non-linear scalar field $\xi$. The rest of fermion fields are usually given large vector-like masses introducing their corresponding left-handed counterparts in extra, incomplete \so5 multiplets. However, this setup breaks gauge invariance because the \so5 multiplets transform non linearly. In fact, it has been shown \cite{illana2022new} that a gauge transformation involves all the \so5 generators, not only those associated to the gauge group, which implies that any \so5 multiplet should be complete because in general a gauge transformation will mix all its components. As a consequence, all the right-handed fermions transforming in a \so5 representation must have the same mass. On the other hand, the singlet fermion field can be given either T-parity assignment: some authors consider the T-even option \cite{delaguilaLeptonFlavorChanging2017,delaguilaInverseSeesawNeutrino2019} while others choose it T-odd \cite{chengLittleHierarchyLittle2004,lowParityLittlestHiggs2004,Reuter:2013iya}. However, only the T-even realization is compatible with gauge invariance \cite{illana2022new}.

The new Littlest Higgs model with T-parity (NLHT) is a minimal extension of the LHT that fulfills the gauge invariance requirements \cite{illana2022new}. To that end, the LHT global symmetry group is minimally enlarged with an extra $\left[\su2\times \u1\right]''_1\times \left[\su2\times \u1\right]''_2$ spontaneously broken to $\left[\su2\times \u1\right]''$ at the same scale $f$ due to the \vev\ of another symmetric tensor. The gauge group is still $\left[\su2\times \u1\right]_1\times \left[\su2\times \u1\right]_2$ preserving the number of gauge bosons. As a consequence, the model includes new particles: a T-odd real triplet and a T-odd singlet in the scalar sector and a doublet of T-even mirror partners and a T-odd singlet in the fermionic sector. All the new fermions and the usual T-odd mirror partners and T-even singlet receive masses proportional to the scale $f$ through a Yukawa Lagrangian. 

In this work we explore the available parameter space compatible with EWPD and cosmological constraints in a simplified version of the model in which the Yukawa couplings of heavy leptons and heavy quarks are mass degenerate. In particular all the new quarks must be heavier than 2 TeV to evade vector-like quark constraints \cite{Vatsyayan:2020jan} while leptons could be lighter. On the other hand, the new scalars are naturally light independent of $f$. Their masses parametrically depend on the Yukawa couplings of the top quark and the heavy fermions. However, they could take any value by tuning these couplings. Using a naturalness argument and choosing the usual heavy photon $A_H$ as a viable candidate for dark matter, we establish an upper and a lower bounds for their masses. The paper is organized as follows. Section~\ref{review} presents a review of the NLHT to define the framework and fix the notation. In section~\ref{Pheno} we study the parameter space and the particle spectrum, paying special attention to the decay channels and lifetime of the new T-odd scalars. The last section is devoted to our conclusions.

%----------------------------------------------------------------------%
%-------------- 	NLHT REVIEW -----------------------------------------%
%----------------------------------------------------------------------%
\section{The new Littlest Higgs model with T-parity}\label{review}

%----------------------------------------------------------------------%
%-------------- 	Global symmetry -------------------------------------%
%----------------------------------------------------------------------%
\subsection{Global symmetries \label{subsec:globalsymmetries}}

Let us briefly review the new Littlest Higgs model with T-parity (NLHT), following closely the notation introduced in ref.~\cite{illana2022new}. 
The model is based on the symmetric coset 
\eq{\label{globalgroup}
G=\su5\times\left[\su2\times \u1\right]''_1\times \left[\su2\times \u1\right]''_2\xrightarrow{\Sigma_0,\widehat{\Sigma}_0}
H=\so5\times \left[\su2\times \u1\right]''
}
parametrized by the vacuum expectation value of two symmetric tensors, 
\eq{
\Sigma_{0}=\left(\begin{array}{ccc}
0_{2\times 2} & 0 & {\bf{1}}_{2\times 2}\\
0 & 1 & 0\\
{\bf{1}}_{2\times2} & 0 & 0_{2\times2}
\end{array}\right),\quad \widehat{\Sigma}_0=\Sigma_0,
\label{Sigma0}
}
where $\Sigma_0$ breaks spontaneously $\su5$ into $\so5$ leaving $14$ Goldstone bosons, while $\widehat{\Sigma}_0$ breaks $\left[\su2\times \u1\right]''_1\times \left[\su2\times \u1\right]''_2$ to $\left[\su2\times \u1\right]''$ leaving $4$ extra Goldstone bosons at the same energy scale $f$, the scale of new physics (NP). On the other hand, one needs to extend the global symmetry with external $\u1'''_1\times \u1'''_2$ in order to accommodate the proper hypercharges for the SM right-handed charged leptons and all quarks (see for instance ref.~\cite{gotoTauMuonLepton2011}).
The unbroken generators preserve the vacuum satisfying the relations
\eq{\label{unbroken}
T^a\Sigma_0+\Sigma_0 T^{a T}&=0,\\
\widehat{T}^a\Sigma_0+\Sigma_0 \widehat{T}^{a T}&=0,
}
where the first set of generators corresponds to $\su5$ and the second to $\left(\left[\su2\times\u1\right]''\right)^2$. The set of broken generators are orthogonal to the unbroken ones satisfying
\eq{
X^a\Sigma_0-\Sigma_0 X^{a T}&=0,\label{brokensu5}\\
\widehat{X}^a\Sigma_0-\Sigma_0 \widehat{X}^{a T}&=0\label{brokensu2u1squared}.
}
The broken generators span two Goldstone matrices $\Pi=\pi^a X^a$ and $\widehat{\Pi}=\widehat{\pi}^a \widehat{X}^{a}$. This allows the introduction of the non-linear fields $\xi$ and $\widehat{\xi}$ that transform under the global symmetry group,
\begin{alignat}{3}\label{xitransformation}
  \xi&= e^{i\Pi/f},\qquad \xi && \xrightarrow{G} V\xi U^{\dagger}=U\xi\Sigma_0 V^T\Sigma_0,\\
  \widehat{\xi}&= e^{i\widehat{\Pi}/f},\qquad \widehat{\xi} && \xrightarrow{G} \widehat{V}\widehat{\xi} \widehat{U}^{\dagger}=\widehat{U}\widehat{\xi}\Sigma_0 \widehat{V}^T\Sigma_0,\label{xihattransformation}
\end{alignat}
where $V$ is a transformation of $\su5$, $U=U\left(V,\Pi\right)$ is the compensating $\so5$ non-linear transformation and $\widehat{V}$ is a transformation of $\left[\su2\times\u1\right]''_1\times \left[\su2\times\u1\right]''_2$ with $\widehat{U}=\widehat{U}(
\widehat{V},\widehat{\Pi})$ being the corresponding $\left[\su2\times\u1\right]''$ non linear transformation.

Note that the second identities in eqs.~(\ref{xitransformation}) and (\ref{xihattransformation}) are derived from the invariance of the symmetric tensor $\Sigma_0$ under the unbroken subgroup \so5, that implies $U\Sigma_0 = \Sigma_0 U^*$, together with the characterization of the broken generators, $X^a\Sigma_0=\Sigma_0X^{aT}$, that implies $\xi^T=\Sigma_0\xi\Sigma_0$. Substituting both relations into the result of the transformation under the global symmetry group, $\xi\to V\xi U^\dagger$ (CCWZ formalism \cite{callanStructurePhenomenologicalLagrangians1969}), we have
$
\xi = \Sigma_0\xi^T\Sigma_0 \to \Sigma_0(V\xi U^\dagger)^T\Sigma_0
    = \Sigma_0 U^*\xi^T V^T\Sigma_0
    = U \Sigma_0 \Sigma_0 \xi \Sigma_0 V^T \Sigma_0
    = U \xi \Sigma_0 V^T \Sigma_0
$
(and similarly for the hatted fields), as we wanted to prove.

We also define two tensor fields $\Sigma$, $\widehat{\Sigma}$ that transform linearly under the global symmetry group,
\begin{alignat}{3}
\Sigma &=\xi\Sigma_0\xi^{T}=\xi^2\Sigma_0,\qquad \Sigma && \xrightarrow{G} V\Sigma V^{T},\label{sigmatransformation}\\
\widehat{\Sigma} &=\widehat{\xi}\Sigma_0\widehat{\xi}^{T}=\widehat{\xi}^2\Sigma_0,\qquad \widehat{\Sigma} && \xrightarrow{G} \widehat{V}\Sigma \widehat{V}^{T}\label{sigmahattransformation}.
\end{alignat}

%----------------------------------------------------------------------%
%--------------- 	Gauge group -----------------------------------------%
%----------------------------------------------------------------------%
\subsection{Gauge group}

A subgroup $G_g=\left[\su2\times \u1\right]_1\times \left[\su2\times \u1\right]_2$ of the full global group $G$ is gauged. It is spanned by the Hermitian and traceless generators
\eq{
Q_{1}^{a}&=\frac{1}{2}\left(\begin{array}{ccc}
\sigma^{a} & 0 & 0\\
0 & 0 & 0\\
0 & 0 & 0_{2\times2}
\end{array}\right) 
\hspace{-2cm} &&,\quad Y_1= \frac{1}{10}\textrm{diag}\left(3,3,-2,-2,-2\right),
\label{generators1}
\\
Q_{2}^{a}&=\frac{1}{2}\left(\begin{array}{ccc}
0_{2\times 2} & 0 & 0\\
0 & 0 & 0\\
0 & 0 & -\sigma^{a*}
\end{array}\right)
\hspace{-2cm} &&,\quad Y_2= \frac{1}{10}\textrm{diag}\left(2,2,2,-3,-3\right),
\label{generators2}
}
with $\sigma^a$ the three Pauli matrices.\footnote{This is possible because we have taken the form of the generators in $\left[\su{2}\times\u{1}\right]''_1\times \left[\su{2}\times\u{1}\right]''_2$ the same as those which span the subgroup $\left[\su{2}\times\u{1}\right]'_1\times \left[\su{2}\times\u{1}\right]'_2 \subset \su5$.} A useful property that the gauge generators satisfy is 
\eq{\label{gaugegeneratorsproperty}
Q^a_1=-\Sigma_0 Q^{a T}_2\Sigma_0,\quad Y_1=-\Sigma_0 Y_2^T\Sigma_0,
}
which relates the generators of both gauge subgroups. The \vev\ along the direction of $\Sigma_0$ and $\widehat{\Sigma}_0=\Sigma_0$ also breaks spontaneously the gauge group down to the diagonal subgroup $\su{2}_L\times\u{1}_Y$ identified as the SM gauge group, generated by the combinations $\left\{Q_1^a+Q_2^a,Y_1+Y_2\right\}\subset \left\{T^a, \widehat{T}^a\right\}$. The orthogonal combination are a subset of the broken generators, $\{Q_1^a-Q_2^a,Y_1-Y_2\}\subset \{X^a, \widehat{X}^a\}$.

The set of broken generators of $\su5$, $\{X^a\}$, spans the first Goldstone matrix,
\eq{
\Pi=\left(\begin{array}{ccccc}-\disp\frac{\omega^0}{2}-\frac{\eta}{\sqrt{20}} & -\disp\frac{\omega^+}{\sqrt{2}} & -i\disp\frac{\pi^+}{\sqrt{2}} & -i\Phi^{++} & -i\disp\frac{\Phi^+}{\sqrt{2}} \\
-\disp\frac{\omega^-}{\sqrt{2}} & \disp\frac{\omega^0}{2}-\frac{\eta}{\sqrt{20}} & \disp\frac{v+h+i\pi^0}{2} & -i\disp\frac{\Phi^+}{\sqrt{2}} & \disp\frac{-i\Phi^0+\Phi^P}{\sqrt{2}} \\
i\disp\frac{\pi^-}{\sqrt{2}} & \disp\frac{v+h-i\pi^0}{2} & \sqrt{\disp\frac{4}{5}}\eta & -i\disp\frac{\pi^+}{\sqrt{2}} &  \disp\frac{v+h+i\pi^0}{2} \\
i\Phi^{--} & i\disp\frac{\Phi^-}{\sqrt{2}} & i\disp\frac{\pi^-}{\sqrt{2}} & -\disp\frac{\omega^0}{2}-\frac{\eta}{\sqrt{20}} & -\disp\frac{\omega^-}{\sqrt{2}} \\
i\disp\frac{\Phi^-}{\sqrt{2}} & \disp\frac{i\Phi^0+\Phi^P}{\sqrt{2}} &  \disp\frac{v+h-i\pi^0}{2} & -\disp\frac{\omega^+}{\sqrt{2}} & \disp\frac{\omega^0}{2}-\frac{\eta}{\sqrt{20}}
\end{array}\right),
}
where $v$ is the Higgs \vev. These Goldstone fields are only charged under $\su5$. Under the SM gauge group this Goldstone matrix includes a complex symmetric $\su{2}$ triplet with hypercharge $Y=1$ and its hermitian conjugate,
\eq{
\Phi=\left(\begin{array}{cc}
-i\Phi^{++} & -i\disp\frac{\Phi^{+}}{\sqrt{2}}\\
-i\disp\frac{\Phi^{+}}{\sqrt{2}} & \disp\frac{-i\Phi^{0}+\Phi^{P}}{\sqrt{2}}
\end{array}\right),\quad \Phi^{\dagger}=\left(\begin{array}{cc}
i\Phi^{--} & i\disp\frac{\Phi^{-}}{\sqrt{2}}\\
i\disp\frac{\Phi^{-}}{\sqrt{2}} & \disp\frac{i\Phi^{0}+\Phi^{P}}{\sqrt{2}}
\end{array}\right),
}
the SM Higgs doublet plus an $\su{2}$ triplet with zero hypercharge,
\eq{
\label{Hdoublet}
H=\left(\begin{array}{c}
i\pi^{+}\\
\disp\frac{v+h+i\pi^{0}}{\sqrt{2}}
\end{array}\right),\quad \omega=\left(\begin{array}{cc}
-\disp\frac{\omega^0}{2} & -\disp\frac{\omega^+}{\sqrt{2}}\\
-\disp\frac{\omega^-}{\sqrt{2}} & \disp\frac{\omega^0}{2}
\end{array}\right)
}
and a singlet, $\eta$. The set of broken generators of $\left[\su2\times \u1\right]''_1\times \left[\su2\times \u1\right]''_2$, denoted $\{\widehat{X}^a\}$, spans the second Goldstone matrix, 
\eq{
\widehat{\Pi}=\left(\begin{array}{ccccc}-\disp\frac{\widehat{\omega}^0}{2}-\frac{\widehat{\eta}}{\sqrt{20}} & -\disp\frac{\widehat{\omega}^+}{\sqrt{2}} & 0 & 0 & 0 \\
-\disp\frac{\widehat{\omega}^-}{\sqrt{2}} & \disp\frac{\widehat{\omega}^0}{2}-\frac{\widehat{\eta}}{\sqrt{20}} & 0 & 0 & 0 \\
0 & 0 & \sqrt{\disp\frac{4}{5}}\widehat{\eta} & 0 &  0 \\
0 & 0 & 0 & -\disp\frac{\widehat{\omega}^0}{2}-\frac{\widehat{\eta}}{\sqrt{20}} & -\disp\frac{\widehat{\omega}^-}{\sqrt{2}} \\
0 & 0 & 0 & -\disp\frac{\widehat{\omega}^+}{\sqrt{2}} & \disp\frac{\widehat{\omega}^0}{2}-\frac{\widehat{\eta}}{\sqrt{20}}
\end{array}\right).
}
These Goldstone fields are charged only under $\left[\su2\times \u1\right]''_1\times \left[\su2\times \u1\right]''_2$. $\widehat{\Pi}$ includes a new SU(2) triplet with zero hypercharge,
\eq{\widehat{\omega}=\left(\begin{array}{cc}
-\disp\frac{\widehat{\omega}^0}{2} & -\disp\frac{\widehat{\omega}^+}{\sqrt{2}}\\
-\disp\frac{\widehat{\omega}^-}{\sqrt{2}} & \disp\frac{\widehat{\omega}^0}{2}
\end{array}\right)
} 
and a singlet, $\widehat{\eta}$. A combination of the scalar fields in $\Pi$ and $\widehat{\Pi}$ with the same T-parity and quantum numbers under the gauged subgroup will become the longitudinal modes of the heavy T-odd gauge bosons. The other combination will remain in the physical spectrum.

%----------------------------------------------------------------------%
%--------------- 	Lagrangian ------------------------------------------%
%----------------------------------------------------------------------%
\subsection{Lagrangian}

%----------------------------------------------------------------------%
%--------------- 	Gauge sector ----------------------------------------%
%----------------------------------------------------------------------%
\subsubsection{Gauge sector}

In the construction of the Lagrangian we take into account the action of the discrete T-parity symmetry introduced to keep the SM gauge bosons T-even and light while the new ones are T-odd and heavy. The action of T-parity consists in an interchange of the two gauge groups,
\eq{
G_1\overset{\rm T}{\longleftrightarrow} G_2,
}
where $G_1=\left[\su{2}\times\u{1}\right]_1$ and $G_2=\left[\su{2}\times\u{1}\right]_2$. This requires that the coupling constants of both copies must be the same $g_1=g_2=\sqrt{2}g$, $g'_1=g'_2=\sqrt{2}g'$, with the first set of couplings referring to \su{2} and the second to \u{1}. The gauge Lagrangian takes the usual form,
\eq{
\Lag_{G}=\sum_{j=1}^2\left[-\frac{1}{2}\textrm{tr}\left(\widetilde{W}_{j\mu\nu}\widetilde{W}_j^{\mu\nu}\right)-\frac{1}{4}B_{j\mu\nu}B_j^{\mu\nu}\right],
\label{lagG}
}
in terms of fields and field strength tensors,
\eq{
\widetilde{W}_{j\mu}=W^a_{j\mu}Q^a_j,\quad \widetilde{W}_{j\mu\nu}=\partial_{\mu}\widetilde{W}_{j\nu}-\partial_{\nu}\widetilde{W}_{j\mu}-i\sqrt{2}g\left[\widetilde{W}_{j\mu},\widetilde{W}_{j\nu}\right],\quad B_{j\mu\nu}=\partial_{\mu}B_{j\nu}-\partial_{\nu}B_{j\mu},
}
where in the first expression the index $j$ is fixed. Before the electroweak SSB, the SM gauge bosons come from the T-even combinations
\eq{
W^{\pm}=\frac{1}{2}\left[\left(W^1_1+W^1_2\right)\mp i\left(W^2_1+W^2_2\right)\right],\quad W^3=\frac{W^3_1+W^3_2}{\sqrt{2}},\quad B=\frac{B_1+B_2}{\sqrt{2}},
\label{gaugeTeven}
}
while the remaining T-odd combinations will define the heavy fields
\eq{
W_H^{\pm}=\frac{1}{2}\left[\left(W^1_1-W^1_2\right)\mp i\left(W^2_1-W^2_2\right)\right],\quad W_H^3=\frac{W^3_1-W^3_2}{\sqrt{2}},\quad B_H=\frac{B_1-B_2}{\sqrt{2}}.
\label{gaugeTodd}
}

%----------------------------------------------------------------------%
%--------------- 	Scalar sector ---------------------------------------%
%----------------------------------------------------------------------%
\subsubsection{Scalar sector}

In order to assign a T-even parity to the SM Higgs field and T-odd parities to the rest of the scalar fields, one defines
\eq{
\Pi&\xrightarrow{\textrm{T}} -\Omega \Pi \Omega, \quad \Omega=\textrm{diag}\left(-1,-1,1,-1,-1\right),\\
\widehat{\Pi}&\xrightarrow{\textrm{T}}-\widehat{\Pi}.
}
It is important to remark that $\Omega$ is an element of the center of the gauge subgroup and consequently commutes with the gauge generators. This fact is directly related to the gauge invariance of the model, as pointed out in ref.~\cite{illana2022new}. The T-parity transformation of the Goldstone matrices defined above reads
\eq{
\xi&\xrightarrow{\textrm{T}}\Omega\xi^{\dagger}\Omega,\quad \Sigma\xrightarrow{\textrm{T}}\widetilde{\Sigma}\equiv\Omega\Sigma_0\Sigma^{\dagger}\Sigma_0\Omega,\\
\widehat{\xi}&\xrightarrow{\textrm{T}}\widehat{\xi}^{\dagger}.
} 
With these ingredients one builds the scalar Lagrangian which is gauge and T-parity invariant,\footnote{A term that mixes both scalar sectors is allowed by gauge invariance. However it would lead to a quadratically divergent contribution to the Higgs mass \cite{illana2022new}.} 
\eq{
\Lag_{S}=\frac{f^2}{8}\textrm{tr}\left[\left(D^{\mu}\Sigma\right)^{\dagger}D_{\mu}\Sigma\right]+\frac{f^2}{8}\textrm{tr}\left[\left(D^{\mu}\widehat{\Sigma}\right)^{\dagger}D_{\mu}\widehat{\Sigma}\right],
\label{lagS}
}
where the covariant derivative is defined as
\eq{\label{scalarderivative}
D_{\mu}\Sigma=\partial_{\mu}\Sigma-\sqrt{2}i\sum_{j=1}^2\left[g W^a_{j\mu}\left(Q^a_j\Sigma+\Sigma Q^{a T}_j\right)-g'B_{j\mu}\left(Y_j\Sigma+\Sigma Y_j^T\right)\right]
}
and the same for $\widehat{\Sigma}$.

%----------------------------------------------------------------------%
%--------------- 	Fermion sector ------------------------------------%
%----------------------------------------------------------------------%
\subsubsection{Fermion sector}

Here we will focus on the quark sector of the theory. Leptons were already described in ref.~\cite{illana2022new}.
First of all, one introduces two left-handed SU(5) quintuplets in the anti-fundamental and fundamental representations, respectively, 
\eq{
\Psi^q_{1}=\left(\begin{array}{c}
-i\sigma^{2}q_{1L}\\
i\chi^q_{1L}\\
-i\sigma^2\widetilde{q}^c_{1L}
\end{array}\right),\quad \Psi^q_{2}=\left(\begin{array}{c}
-i\sigma^2\widetilde{q}^c_{2L}\\
i\chi^q_{2L}\\
-i\sigma^{2}q_{2L}
\end{array}\right).
\label{newsu5multiplets}
}
These \su5 multiplets must be complete to ensure gauge invariant mass terms. Under a gauge transformation,
\eq{
\Psi_1^{q}\xrightarrow{G_g} V_g^{*}\Psi^{q}_1,\quad \Psi^{q}_2\xrightarrow{G_g} V_g\Psi^{q}_2,
}
while under T-parity,
\eq{
\quad\Psi^{q}_1\xrightarrow{\textrm{T}}\Omega\Sigma_0\Psi^{q}_2.\label{Tevenfermions}
}
Consequently, one can define T-even and T-odd combinations given respectively by
\eq{
\quad&\Psi^{q}_{+}=\frac{\Psi^{q}_1+\Omega\Sigma_0\Psi^{q}_2}{\sqrt{2}},\quad \Psi^{q}_{-}=\frac{\Psi^{q}_1-\Omega\Sigma_0\Psi^{q}_2}{\sqrt{2}}.
}
Except for the T-even combination of SM quark doublets,
\eq{
q_{L}=\frac{q_{1L}-q_{2L}}{\sqrt{2}},
}
the rest needs to be paired with a right-handed fermion to get a vector-like mass. To that end, an SO(5) and an $\left[\su2\times\u1\right]''$ right-handed quintuplet are introduced,
\eq{
\Psi^{q}_{R}=\left(\begin{array}{c}
-i\sigma^{2}(\widetilde{q}^{c}_{-})_{R}\\
i\left(\chi^q_{+}\right)_{R}\\
-i\sigma^{2}q_{HR}
\end{array}\right),\quad  \widehat{\Psi}^{q}_{R}=\left(\begin{array}{c}
-i\sigma^{2}(\widetilde{q}^{c}_{+})_{R}\\
i\left(\chi^q_{-}\right)_{R}\\
0_2
\end{array}\right).
\label{psiR}
}
We denote with a subscript $\pm$ the T-parity assignment of the fermion fields except for the SM fermions, which are T-even, and the `mirror' quarks $q_H$, which are T-odd.
The doublet $(\widetilde{q}^c_{\pm})_R$ describes the `mirror partner' quarks and $\left(\chi^q_{\pm}\right)_{R}$ are \su2 singlets. The transformation under the gauged subgroup reads
\eq{\label{PsiRtransformation}
\Psi^{q}_R \xrightarrow{G_g} U_g\Psi^{q}_R,\quad \widehat{\Psi}^{q}_R  \xrightarrow{G_g} \widehat{U}_g  \widehat{\Psi}^{q}_R,
}
where $U_g(V_g,\Pi)$ and $\widehat{U}_g(V_g,\widehat{\Pi})$ are in general non linear transformations of SO(5) and $\left[\su2\times \u1\right]''$, respectively, restricted to the gauged subgroup for a given $V_g$.  
In fact, according to the CCWZ formalism, for any transformation $V\in G$ (the global symmetry group) there is an $U\in H$ (the unbroken subgroup of $G$) such that $\xi\to V\xi U^{-1}$. If in particular $V\in H$ then $U=V$, independent of the Goldstone fields, so in this case $\xi$ transforms linearly. However if $V\notin H$ then $U=U(V,\Pi)$ and the transformation is non linear. Therefore if we restrict ourselves to gauge transformations $V_g\in G_g\subset G$ the corresponding $U_g$ is non linear unless $V_g$ is a Standard Model gauge transformation, because then $V_g\in H$.

The T-parity transformation of these right-handed multiplets is given by
\eq{
\label{TevenPsiRtransformation}\Psi^{q}_R\xrightarrow{\textrm{T}}\Omega\Psi^{q}_R,\quad \widehat{\Psi}^{q}_R\xrightarrow{\textrm{T}}-\Omega\widehat{\Psi}^{q}_R,
}
in order to assign the proper T-parity to the quarks. With this in mind, one can construct the two needed Yukawa Lagrangians to give masses to the heavy quarks,
\eq{
\Lag^{q}_{Y_H}&=-\kappa_{q} f\left(\overline{\Psi}^{q}_2\xi+\overline{\Psi}^{q}_1\Sigma_0\xi^{\dagger}\right)\Psi^{q}_R+\hc,\label{lagkappaa}\\
\Lag^{q}_{\widehat{Y}_H}&=-\widehat{\kappa}_{q} f\left(\overline{\Psi}^{q}_2\widehat{\xi}-\overline{\Psi}^{q}_1\Sigma_0\widehat{\xi}^{\dagger}\right)\widehat{\Psi}^{q}_R+\hc,\label{lagkappab}
}
tailored to provide the mirror quarks, the T-even singlet and the T-odd mirror partner quarks with masses of order $\kappa_q f$, and the T-odd singlet and the T-even mirror partner quarks with masses $\widehat{\kappa}_q f$. In this way, only the first set couples to the Higgs field. As one can notice, this construction is similar to that for leptons. This cancels quadratic divergences of the Higgs mass and prevents unwanted infinite contributions to lepton flavor changing Higgs decays as was discussed in refs.~\cite{delaguilaInverseSeesawNeutrino2019} and \cite{illana2022new}. 

The top quark is the heaviest SM fermion and the only contributing significantly to the quadratic divergences of the Higgs mass at the loop level. To cancel these corrections, in the LHT as well as in the NLHT, one introduces a set of `cancelon' fields: top partners $T_{1L}$, $T_{2L}$ and their corresponding right-handed counterparts $T_{1R}$ and $T_{2R}$. In order to provide masses to the top quark and its partners while avoiding quadratic divergences to the Higgs mass, the following Lagrangian inspired in \cite{Han:2005ru,hubiszPhenomenologyLittlestHiggs2005} can be introduced,
\eq{
\Lag^t_{Y}=&-i\frac{\lambda_1 f}{4}\epsilon_{ijk}\epsilon_{xy}\left[\left(\overline{Q}^t_1\right)_i\Sigma_{jx}\Sigma_{ky}+\left(\overline{Q}^t_2\Sigma_0\Omega\right)_i\widetilde{\Sigma}_{jx}\widetilde{\Sigma}_{ky}\right]t_R\nn\\
&-\frac{\lambda_2 f}{\sqrt{2}}\left(\overline{T}_{1L}\widehat{X}T_{1R}+\overline{T}_{2L}\widehat{X}^{*}T_{2R}\right)+\hc\label{toplag},
}
where $\left\{i,j,k\right\}=1,2,3$ and $\left\{x,y\right\}=4,5$ and the \su2 singlet $t_R$ is the T-even right-handed top quark. 
The left-handed quarks are embedded in the incomplete \su5 multiplets
\eq{\label{PsiQ}
Q_1^{t}=\left(\begin{array}{c}
-i\sigma^2 \mathcal{T}_{1L}\\
i\dis T_{1L}\\
0_{2}
\end{array}\right),\quad Q_2^{t}=\left(\begin{array}{c}
0_2\\
i\dis T_{2L}\\
-i\sigma^2 \mathcal{T}_{2L}
\end{array}\right)
}
where the SU(2) doublets
\eq{
\mathcal{T}_{rL}=\left(\begin{array}{c}
t_{rL}\\
b_{rL}
\end{array}\right)
}
have the same transformation properties under the gauge group and T-parity as those in $\Psi^q_{1,2}$ of eq.~(\ref{newsu5multiplets}) and $(T_1)_{L,R}\xrightarrow{\textrm{T}} (T_2)_{L,R}$. The main difference with respect to ref.~\cite{Han:2005ru,hubiszPhenomenologyLittlestHiggs2005}, is the presence of the scalar field $\widehat{X}=\widehat{\Sigma}^{-1/2}_{33}=e^{-i\sqrt{\frac{4}{5}}\widehat{\eta}}$ with hypercharges $\left(Y_1,Y_2\right)=\left(\frac{1}{5},-\frac{1}{5}\right)$ under the gauge group, and its complex conjugate $\widehat{X}^*$. This field is introduced in order to change the hypercharges of the right-handed top quark partners $(T_{1,2})_R$ to those of the right-handed $t_R$. These new hypercharges are more natural since all fermions in a linear representation of the global group requiring extra hypercharge receive half of it in each of the external $\u1'''_j$, $j=1,2$ (see Table~\ref{table:quantumnumbersquarkslinear}). The same applies to fermions that transform in a non linear representation (see Table.~\ref{table:quantumnumbersnonlinearq}), which receive the extra hypercharge needed from the diagonal $\u1'''$. This is because the hypercharge under the gauged part of $\su5\times\left(\left[\su2\times\u1\right]''\right)^2$ is fixed by the form of the generators in eqs.~(\ref{generators1}) and (\ref{generators2}).\footnote{Something similar happens to the SM right-handed charged leptons $\ell_R$ and the down type quarks $d_R$. 
They are singlets under $\su5\times\left(\left[\su2\times\u1\right]''\right)^2$, receiving their hypercharges $\left(Y_1,Y_2\right)=\left(-\frac{1}{2},-\frac{1}{2}\right)$ and $\left(-\frac{1}{6},-\frac{1}{6}\right)$, respectively, from the external $\u1_1'''\times \u1_2'''$.}

\begin{table}
\centering
\begin{tabular}{|c|c|c|c|}
\hline
 & $\left[\su{2}'\times \u{1}'\right]^2\subset \textrm{SU(5)}$  &  $\left[\u1''' \right]^2$  & $\left[\su{2}\times \u{1}\right]^2$  \T\B\\ 
\hline
$q_{2L}$ & $\left(1,2\right)_{\left(-\frac{1}{5},-\frac{3}{10}\right)}$ & $\left(\frac{1}{3},\frac{1}{3}\right)$ & $\left(1,2\right)_{\left(\frac{2}{15},\frac{1}{30}\right)}$   \T\B\\ 
$q_{1L}$ & $\left(2,1\right)_{\left(-\frac{3}{10},-\frac{1}{5}\right)}$ & $\left(\frac{1}{3},\frac{1}{3}\right)$ &$\left(2,1\right)_{\left(\frac{1}{30},\frac{2}{15}\right)}$ \T\B\\ 
$\chi^q_{2L}$ & $\left(1,1\right)_{\left(-\frac{1}{5},\frac{1}{5}\right)}$ & $\left(\frac{1}{3},\frac{1}{3}\right)$ &$\left(1,1\right)_{\left(\frac{2}{15},\frac{8}{15}\right)}$ \T\B\\
$\chi^q_{1L}$ & $\left(1,1\right)_{\left(\frac{1}{5},-\frac{1}{5}\right)}$ & $\left(\frac{1}{3},\frac{1}{3}\right)$ &$\left(1,1\right)_{\left(\frac{8}{15},\frac{2}{15}\right)}$   \T\B\\
$\widetilde{q}^c_{2L}$ &  $\left(2,1\right)_{\left(\frac{3}{10},\frac{1}{5}\right)}$  & $\left(\frac{1}{3},\frac{1}{3}\right)$ & $\left(2,1\right)_{\left(\frac{19}{30},\frac{8}{15}\right)}$    \T\B\\
$\widetilde{q}^c_{1L}$ &  $\left(1,2\right)_{\left(\frac{1}{5},\frac{3}{10}\right)}$ & $\left(\frac{1}{3},\frac{1}{3}\right)$ &$\left(1,2\right)_{\left(\frac{8}{15},\frac{19}{30}\right)}$  \T\B\\
$d_{R}$ &  $\left(1,1\right)_{\left(0,0\right)}$  & $\left(-\frac{1}{6},-\frac{1}{6}\right)$ & $\left(1,1\right)_{\left(-\frac{1}{6},-\frac{1}{6}\right)}$  \T\B\\
$u_{R}, T_{2R}, T_{1R}$ &  $\left(1,1\right)_{\left(0,0\right)}$ & $\left(\frac{1}{3},\frac{1}{3}\right)$ & $\left(1,1\right)_{\left(\frac{1}{3},\frac{1}{3}\right)}$  \T\B\\
$T_{2L}$ &  $\left(1,1\right)_{\left(-\frac{1}{5},\frac{1}{5}\right)}$ & $\left(\frac{1}{3},\frac{1}{3}\right)$ &$\left(1,1\right)_{\left(\frac{2}{15},\frac{8}{15}\right)}$  \T\B\\
$T_{1L}$ &  $\left(1,1\right)_{\left(\frac{1}{5},-\frac{1}{5}\right)}$ & $\left(\frac{1}{3},\frac{1}{3}\right)$ &$\left(1,1\right)_{\left(\frac{8}{15},\frac{2}{15}\right)}$  \T\B\\[1ex]
\hline
\end{tabular}
\caption{Charge assignments under the different $\left[\su{2}\right]$'s and $\left[\u{1}\right]$'s of quarks transforming in a linear representation. These fields are singlets under $\left[\su{2}\times \u{1}\right]''$. Note that $\mathcal{T}_{iL}$ and $t_R$ are the third family of $q_{iL}$ and $u_R$, respectively}\label{table:quantumnumbersquarkslinear}
\end{table}

\begin{table}
\centering%\begin{center}
\begin{tabular}{|c|c|c|c|c|}
\hline
 &  $\su{2}'\times \u{1}'\subset \textrm{SO(5)} $  &  $\su{2}''\times \u{1}''$  & $\left[\u1'''\right]$ & $\left[\su{2}\times \u{1}\right]$    \T\B\\ 
\hline
$\left(\begin{array}{c}
-i\sigma^{2}(\widetilde{q}^{c}_{-})_{R}\\
i(\chi^q_{+})_{R}\\
-i\sigma^{2}q_{HR}
\end{array}\right)$ & $\left(\begin{array}{c}
2_{\frac{1}{2}}\\
1_0\\
2_{-\frac{1}{2}}
\end{array}\right)$ & $1_{0}$& $\frac{2}{3}$ & $\left(\begin{array}{c}
2_{\frac{7}{6}}\\
1_\frac{2}{3}\\
2_{\frac{1}{6}}
\end{array}\right)$  \T\B\\
$(\widetilde{q}^c_{+})_R$ &  $1_0$ &  $2_{\frac{1}{2}}$ & $\frac{2}{3}$ &$2_{\frac{7}{6}}$    \T\B\\
$(\chi^q_{-})_{R}$ & $1_0$ & $1_{0}$ & $\frac{2}{3}$  & $1_\frac{2}{3}$   \T\B\\[1ex]
\hline
\end{tabular}
\caption{Charge assignments under the different $\left[\su{2}\right]$'s and $\left[\u{1}\right]$'s of right-handed quarks transforming in a non linear representation.}\label{table:quantumnumbersnonlinearq}
\end{table}

There exist other possible ways to introduce fermions transforming under the same symmetry group that are also free from quadratically divergent contributions to the Higgs mass. For instance, in ref.~\cite{pappadopuloTparityItsProblems2011} they choose to embed the left-handed doublets $q_{1L}$ and $q_{2L}$ in representations of the external $\left[\su2\times\u1\right]''_1\times \left[\su2\times\u1\right]''_2$ while the doublet of right-handed mirror fermions transforms under the diagonal $\left[\su2\times\u1\right]''$. These are coupled through the analog to $\widehat{\xi}$ generating mirror fermion masses of order $\kappa f$. This construction does not require the introduction of mirror partner fermions nor the singlet $\chi$ and is gauge invariant. Since the Higgs belongs to $\xi$, its mass is protected against radiative corrections from this sector. Besides, the global invariance under the external $\left[\su2\times\u1\right]''_1\times \left[\su2\times\u1\right]''_2$ ensures that the new T-odd scalars do not develop a contribution to their mass terms coming from mirror fermion interactions at one loop. However, we prefer to keep the structure of the original Yukawa Lagrangian to give masses of order $\kappa f$ to mirror fermions, compatible with gauge invariance \cite{delaguilaLeptonFlavorChanging2017,delaguilaInverseSeesawNeutrino2019,illana2022new}. 
Regarding the SM fermions, which are the T-even combination of $q_{1L}$ and $q_{2L}$, they are embedded in incomplete SU(5) representations in order to provide them with masses by the same type of Yukawa Lagrangian as in our model.
 
The CCWZ formalism provides us with the kinetic terms and the gauge interactions for all quarks,
\eq{
\Lag^q_F = \Lag^q_{F_L} + \Lag^q_{F_R} + \Lag^q_{\widehat{F}_R}
}
where
\eq{
\Lag^q_{F_L} &=
i\overline{\Psi}^q_1\gamma^{\mu}D^{q*}_{\mu}\Psi^q_1+i\overline{\Psi}^q_2\gamma^{\mu}D^q_{\mu}\Psi^q_2 
\label{LagFL}
\\
\Lag^q_{F_R} &=
i\overline{\Psi}^q_R\gamma^{\mu}\left[\partial_{\mu}+\frac{1}{2}\xi^{\dag}\left(D^q_{\mu}\xi\right)+\frac{1}{2}\xi\Sigma_0D^{q*}_{\mu}\left(\Sigma_0\xi^{\dag}\right)\right]\Psi^q_R,\\
\Lag_{\widehat{F}_R}^{q} &=
i\overline{\widehat{\Psi}}^q_R\gamma^{\mu}\left[\partial_{\mu}+\frac{1}{2}\widehat{\xi}^{\dag}\left(D^q_{\mu}\widehat{\xi}\right)+\frac{1}{2}\widehat{\xi}\Sigma_0 D^{q*}_{\mu}\left(\Sigma_0\widehat{\xi}^{\dag}\right)\right]\widehat{\Psi}^q_R,
}
with the covariant derivatives
\eq{
  D^q_{\mu}=\partial_{\mu}&-\sqrt{2}i g \left(W^a_{1\mu} Q^a_1+W^a_{2\mu} Q^a_2\right)\nn\\&+\sqrt{2}i g' \left[B_{1\mu}\left(Y_1+\frac{1}{3}\mathbb{1}_{5\times 5}\right)+B_{2\mu} \left(Y_2+\frac{1}{3}\mathbb{1}_{5\times 5}\right)\right],\\
  D^{q*}_{\mu}=\partial_{\mu}&+\sqrt{2}i g \left(W^a_{1\mu} Q^{a T}_1+W^a_{2\mu} Q^{a T}_2\right)\nn\\&-\sqrt{2}i g' \left[B_{1\mu}\left(Y_1-\frac{1}{3}\mathbb{1}_{5\times 5}\right)+B_{2\mu} \left(Y_2-\frac{1}{3}\mathbb{1}_{5\times 5}\right)\right],
  \label{covderfer}
}
where we have used the properties of gauge generators in eq.~(\ref{gaugegeneratorsproperty}) and we have taken into account the extra hypercharges of quarks showed in Tables~\ref{table:quantumnumbersquarkslinear} and \ref{table:quantumnumbersnonlinearq}. 

Finally, the kinetic term and gauge interactions for the SM right-handed top quark and its corresponding top partners are given by 
\eq{\label{LagFprime}
\Lag^t_{F'}&=i\overline{t}_R\left(\partial_{\mu}-i\frac{2}{3} g'B_{\mu}\right)t_{R},\\
\Lag^{T_1,T_2}_{F'}&=i\overline{T}_{1L}\left[\partial_{\mu}-\sqrt{2}i g'\left(\frac{8}{15}B_{1\mu}+\frac{2}{15}B_{2\mu}\right)\right]T_{1L}+i\overline{T}_{1R}\left(\partial_{\mu}-\frac{2}{3}i g'B_{\mu}\right)T_{1R} \nn\\
&+i\overline{T}_{2L}\left[\partial_{\mu}-\sqrt{2}i g'\left(\frac{2}{15}B_{1\mu}+\frac{8}{15}B_{2\mu}\right)\right]T_{2L}+i\overline{T}_{2R}\left(\partial_{\mu}-\frac{2}{3}i g'B_{\mu}\right)T_{2R}.
}

%----------------------------------------------------------------------%
%--------------- 	Mass eigenfields ------------------------------------%
%----------------------------------------------------------------------%
\subsection{Physical particles}

%----------------------------------------------------------------------%
%---------------  Gauge bosons ----------------------------------------%
%----------------------------------------------------------------------%

\subsubsection{Gauge bosons}

After the electroweak SSB, the SM gauge bosons are obtained from the T-even fields of eq.~(\ref{gaugeTeven}) by diagonalizing the Lagrangian $\Lag_S$ of eq.~(\ref{lagS}),
\eq{
W^{\pm}=\frac{1}{\sqrt{2}}\left(W^1\mp i W^2\right),\quad \left(\begin{array}{c}
Z\\
A
\end{array}\right)=\left(\begin{array}{cc}
c_{W} & s_{W}\\
-s_{W} & c_{W}
\end{array}\right)\left(\begin{array}{c}
W^{3}\\
B
\end{array}\right)
}
with
\eq{
W^a=\frac{W^a_1+W^a_2}{\sqrt{2}},\quad \quad B=\frac{B_1+B_2}{\sqrt{2}}.
}
To get the physical T-odd gauge bosons, one needs to expand $\Lag_S$ up to order $v^2/f^2$ to derive the heavy physical fields from those in eq.~(\ref{gaugeTodd}),
\eq{
W^{\pm}_H=\frac{1}{\sqrt{2}}\left(W^1_H\mp i W^2_H\right),\quad \left(\begin{array}{c}
Z_{H}\\
A_{H}
\end{array}\right)=\left(\begin{array}{cc}
1 & -x_{H}\frac{v^{2}}{f^{2}}\\
x_{H}\frac{v^{2}}{f^{2}} & 1
\end{array}\right)\left(\begin{array}{c}
W_{H}^{3}\\
B_{H}
\end{array}\right)
}
with
\eq{
W^a_H=\frac{W^a_1-W^a_2}{\sqrt{2}},\quad B_H=\frac{B_1-B_2}{\sqrt{2}},\quad x_H=\frac{5 g g'}{8\left(5g^2-g'^2\right)}.
}
Their corresponding masses to order $v^2/f^2$ are
\eq{\label{gaugebosonmasses}
& M_W=\frac{g v}{2}\left(1-\frac{v^2}{12f^2}\right),\quad M_Z=M_W/c_W,\quad e= g s_W= g' c_W, \nn\\
& M_{W_H}=M_{Z_H}=\sqrt{2}g f\left(1-\frac{v^2}{16f^2}\right),\quad M_{A_H}=\sqrt{\frac{2}{5}}g' f\left(1-\frac{5v^2}{16f^2}\right).
}
Notice that even though the gauge bosons are the same as in the original model, the masses of the T-odd combinations are at leading order a factor of $\sqrt{2}$ larger (see for instance \cite{delaguilaLeptonFlavorChanging2017}). This is because the new extra scalar sector parametrized by $\widehat{\Sigma}$ also takes the \vev\ $\Sigma_0$ hence giving an additional contribution to the heavy gauge boson masses. However the T-even gauge bosons couple only to the Higgs field, belonging to $\Sigma$, so their masses do not change. Let us point out that $A_H$ is always the lightest of the T-odd gauge bosons.

%----------------------------------------------------------------------%
%--------------- Scalars after gauge fixing ---------------------------%
%----------------------------------------------------------------------%
\subsubsection{Scalars after gauge fixing}\label{subsec:scalars}

The kinetic terms of the scalar fields we have introduced are neither diagonal nor canonically normalized after the SSB. Besides, the new T-odd scalars from the extra group mix with old ones with the same quantum numbers. In order to define the physical scalars and identify the actual would-be-Goldstone fields one has to perform some redefinitions \cite{illana2022new}. The redefined scalars $\eta$, $\omega^0$ and $\omega^{\pm}$ are the would-be-Goldstone bosons of the SSB of the gauge group down to the SM gauge group, eaten by $A_H$, $Z_H$ and $W^{\pm}_H$. Similarly, $\pi^0$ and $\pi^{\pm}$ become the would-be-Goldstone bosons of the SSB of the SM gauge group down to $\u{1}_\textrm{em}$, eaten by $Z$ and $W^{\pm}$. The remaining scalar fields are all physical. They are the Higgs boson, the usual T-odd triplet with hypercharge $Y=1$ composed of $\Phi^{\pm\pm}$, $\Phi^{\pm}$, $\Phi^0$ and $\Phi^P$, and the four new T-odd scalars of hypercharge $Y=0$, a singlet $\widehat{\eta}$ and a triplet composed of $\widehat{\omega}^0$ and $\widehat{\omega}^{\pm}$. All of them get a mass by gauge and Yukawa interactions from the Coleman-Weinberg potential. 

The Higgs mass exhibits a dependence on the cut-off scale, that is just logarithmic as a result of the collective symmetry breaking mechanism,
\eq{
\label{hmass}
M^2_h&=
\frac{f^2}{8\pi^2}\log\Lambda^2\left(3\lambda_1^2\lambda_2^2-6g^4-\frac{2}{5}g'^4-12T_\kappa\right),
}
where $T_{\kappa}=\textrm{tr}\left(\kappa_l\kappa_l^{\dagger}\widehat{\kappa}_l\widehat{\kappa}_l^{\dagger}\right)+3 \textrm{tr}\left(\kappa_q\kappa_q^{\dagger}\widehat{\kappa}_q\widehat{\kappa}_q^{\dagger}\right)$ includes both heavy quark and heavy lepton Yukawa couplings \cite{illana2022new}. 
The influence of $T_\kappa$ on the Higgs mass corrections is unavoidable in the NLHT. There is no dependence on $\kappa$'s in the Littlest Higgs model without T-parity because mirror fermions are not needed in that case. The implementation of T-parity requires mirror fermions with masses proportional to $\kappa$. In the LHT the cancellation of quadratic corrections to the Higgs mass is guaranteed if the right-handed quintuplet contains the singlet $\chi$ \cite{illana2022new}, and logarithmic corrections cancel as long as $\chi$ is T-even and has a mass $M_\chi=\sqrt{2}\kappa f$. However, in the NLHT with a complete multiplet $\Psi_R$ the Yukawa Lagrangian $\mathcal{L}_{Y_H}$ is \su5 invariant because the left-handed fermion multiplets $\Psi_{1,2}$ are also complete, so it cannot generate any contribution to scalar masses through loops. On the other hand, $\mathcal{L}_{\widehat{Y}_H}$ shares the left-handed multiplets with $\mathcal{L}_{Y_H}$ but it is not invariant under global transformations of any of the symmetry group factors. As a consequence, the breaking of the global symmetry due to $\mathcal{L}_{\widehat{Y}_H}$, which does not contain the Higgs, can only propagate to the Higgs through $\Psi_{1,2}$, resulting just in a logarithmically divergent contribution to its mass proportional to $\kappa^2\widehat{\kappa}^2$ at one loop, a manifestation of the collective symmetry breaking mechanism.

The rest of the scalar masses are then given in terms of the Higgs mass and \vev\ by

\eq{
M^2_{\Phi}&=2\frac{f^2}{v^2}M_h^2,\label{mphi}\\
M^2_{\widehat{\omega}}&=8M^2_h\,
\frac{g^4+T_{\kappa}}{3\lambda_1^2\lambda_2^2-6g^4-\frac{2}{5}g'^4
-12 T_{\kappa}},
\label{mw}
\\
M^2_{\widehat{\eta}}&=\frac{72}{5}M^2_h\,
\frac{T_{\kappa}}{3\lambda_1^2\lambda_2^2-6g^4-\frac{2}{5}g'^4
-12 T_{\kappa}},
\label{meta}
}
The components of both triplets are mass-degenerate at this order. Notice that the masses of the new T-odd scalars are proportional to the Higgs mass and independent of the scale $f$, and thus naturally light. This is because they inherit part of the symmetry from the would-be Goldstone bosons to be eaten after the SSB at the scale $f$. Similarly to the Higgs mass, they receive contributions proportional to $T_{\kappa}$. On the other hand, the triplet $\widehat\omega$ has zero hypercharge and the gauge couplings can only generate a contribution to its mass proportional to $g^4$, while the singlet $\widehat\eta$ does not receive any contribution from gauge interactions. Numerically, their masses could take any value since they are strictly increasing functions of $T_{\kappa}$ for fixed $\lambda_{1,2}$, keeping in mind that a successful electroweak SSB requires that theHiggs mass in eq.~(\ref{hmass}) remains positive. 
From these equations we find
\eq{\label{metaproptomomega}
M^2_{\widehat{\eta}}=\frac{9}{5}\frac{T_{\kappa}}{T_{\kappa}+g^4}M^2_{\widehat{\omega}}.
}
One may argue that the new coupling between the top quark partners and the scalar singlet $\widehat{\eta}$ in eq.~(\ref{toplag}) would add a correction to $M_{\widehat{\eta}}$ not accounted in ref.~\cite{illana2022new}. However, a simple one loop calculation reveals that this is not the case. Due to the identification $\widehat{X}\orden e^{i\widehat{\eta}}$, the scalar $\widehat{\eta}$ can be understood as the Goldstone boson of the spontaneous breaking of the gauged $\left[\u1\right]^2$ to $\u1$. This prevents the generation of a new contribution to its mass at one loop.

Finally, the Higgs quartic coupling is given by
\eq{\label{higgsquartic}
\lambda=\frac{1}{16\pi^2}\frac{\Lambda^2}{f^2}\left(g^2+g'^2+3\lambda^2_1\right).
}
This expression relates the cutoff scale $\Lambda$ to the top Yukawa coupling $\lambda_1$. The value of this coupling is fixed to $\lambda=\frac{1}{2}(M_h/v)^2$.

%----------------------------------------------------------------------%
%--------------- 	Quark masses --------------------------------------%
%----------------------------------------------------------------------%
\subsubsection{Quark masses}

Once the fermion content of the model is extended beyond one family, the Yukawa couplings must be understood as $3\times3$ matrices in flavor space. Lepton masses and mixings were tackled in ref.~\cite{illana2022new}. Here we will restrict ourselves to the case of quarks. Omitting family indices, for each of the three SM (T-even) left-handed quark doublets $q_L$ there is a vector-like doublet of heavy T-odd mirror quarks $q_H$,
\eq{
&q_L=\left(\begin{array}{c}
u_L\\
d_L
\end{array}\right)=\frac{q_{1L}-q_{2L}}{\sqrt{2}},\quad 
q_{HL}=
\left(\begin{array}{c}
u_{HL}\\
d_{HL}
\end{array}\right)=\frac{q_{1L}+q_{2L}}{\sqrt{2}},\quad 
q_{HR}=\left(\begin{array}{c}
u_{HR}\\
d_{HR}
\end{array}\right),
}
where
\eq{
q_{rL}=\left(\begin{array}{c}
u_{rL}\\
d_{rL}
\end{array}\right), \quad r=1,2
}
are part of the SU(5) multiplets $\Psi^{q}_r$ in eq.~(\ref{newsu5multiplets}) and  $q_{HR}$ is part of the SO(5) multiplets $\Psi^{q}_R$ in eq.~(\ref{psiR}). The SM right-handed quarks $d_R$, $u_R$ are singlets under the full SU(5) but have hypercharges under the external $\u{1}'''$ groups (see Table~\ref{table:quantumnumbersquarkslinear}).

In addition there are two heavy right-handed mirror-partner doublets,
\eq{
(\widetilde{q}^c_{-})_{R}=\left(\begin{array}{c}
(\widetilde{u}^c_{-})_{R}\\
(\widetilde{d}^c_{-})_{R}
\end{array}\right),\quad (\widetilde{q}^c_{+})_{R}=\left(\begin{array}{c}
(\widetilde{u}^c_{+})_{R}\\
(\widetilde{d}^c_{+})_{R}
\end{array}\right).
}
The first one is T-odd and it is necessary to complete the multiplets $\Psi^{q}_R$ together with $(\chi^{q}_+)_R$, while the second is T-even and lays in the incomplete multiplet $\widehat{\Psi}^{q}_R$ in eq.~(\ref{psiR}) along with $(\chi^{q}_{-})_R$. Their corresponding left-handed counterparts come from the SU(5) multiplets, 
\eq{
& (\widetilde{q}^c_{-})_L
=\left(\begin{array}{c}
(\widetilde{u}^c_{-})_{L},\\
(\widetilde{d}^c_{-})_{L}
\end{array}\right)
=\frac{(\widetilde{q}^c_1)_L+(\widetilde{q}^c_2)_L}{\sqrt{2}}
,\quad (\widetilde{q}^c_{+})_L
=\left(\begin{array}{c}
(\widetilde{u}^c_{+})_{L},\\
(\widetilde{d}^c_{+})_{L}
\end{array}\right)
=\frac{(\widetilde{q}^c_1)_L-(\widetilde{q}^c_2)_L}{\sqrt{2}},
}
with
\eq{
(\widetilde{q}^c_r)_L=\left(\begin{array}{c}
(\widetilde{u}^c_r)_L\\
(\widetilde{d}^c_r)_L
\end{array}\right),\quad r=1,2.
}
The \su2 singlets $(\chi^{q}_{+})_R$ and $(\chi^{q}_{-})_R$ complete the set of quarks common to all families. Their left-handed counterparts are the combinations with proper T-parities of the fields $\chi^{q}_{1L}$ and $\chi^{q}_{2L}$ completing the SU(5) multiplets,
\eq{
(\chi^{q}_{+})_L=\frac{\chi^{q}_{1L}+\chi^{q}_{2L}}{\sqrt{2}},\quad (\chi^{q}_{-})_L=\frac{\chi^{q}_{1L}-\chi^{q}_{2L}}{\sqrt{2}}.
}
Finally, the top sector includes a set of top partners $T_{1L}$ and $T_{2L}$, belonging to $Q_1^t $ and $Q_2^t$ respectively, and their right-handed counterparts $T_{1R}$, $T_{2R}$. Their corresponding T-parity eigenstates are
\eq{
\left(T_{+}\right)_{L,R}=\frac{\left(T_{1}\right)_{L,R}+\left(T_{2}\right)_{L,R}}{\sqrt{2}},\quad \left(T_{-}\right)_{L,R}=\frac{\left(T_{1}\right)_{L,R}-\left(T_{2}\right)_{L,R}}{\sqrt{2}}.
}

In the following we present the masses for all quarks at leading order. For simplicity we will assume flavor diagonal and degenerate Yukawa couplings in the rest of this work. The mirror quarks $q_H$, the T-even \su2 singlet $\chi_+^q$ and the T-odd mirror partner quarks $\widetilde{q}^c_{-}$ receive degenerate masses from the Lagrangian in eq.~(\ref{lagkappaa}) of size
\eq{
m_{d_H}=\sqrt{2}\kappa_{q} f
}
while the T-odd \su2 singlet $\chi_{-}^q$ and the T-even mirror partner quarks $\widetilde{q}^c_{+}$ receive masses from the Lagrangian in eq.~(\ref{lagkappab}),
\eq{
m_{\widetilde{q}^c_+}=\sqrt{2}\widehat{\kappa}_q f.
}

Finally, the top quark and its corresponding T-even partner mix in eq.~(\ref{toplag}). After rotating the fields at leading order we have
\eq{\label{eq:toprotation}
\left(\begin{array}{c}
t_R\\
\left(T_{+}\right)_R
\end{array}\right)\rightarrow  \left(\begin{array}{cc}
c_R & s_R\\
-s_R & c_R
\end{array}\right)\left(\begin{array}{c}
t_R\\
\left(T_{+}\right)_R
\end{array}\right),\quad c_R=\frac{\lambda_2}{\sqrt{\lambda_1^2+\lambda_2^2}},\quad s_R=\frac{\lambda_1}{\sqrt{\lambda_1^2+\lambda_2^2}},
}
the physical mass of the top quark in terms of $\lambda_{1,2}$ reads
\eq{\label{topmass}
m_t=\frac{1}{\sqrt{2}}\frac{\lambda_1 \lambda_2 }{\sqrt{\lambda_1^2+\lambda_2^2}}v,
}
and its T-even and T-odd partners have masses
\eq{\label{toppartnermasses}
M_{T_{+}}=\sqrt{\frac{\lambda_1^2+\lambda_2^2}{2}}f, \quad 
M_{T_{-}}=\frac{\lambda_2}{\sqrt{2}}f.
}
Notice that the T-even top partner is always heavier than the T-odd one.

%----------------------------------------------------------------------%
%-------------- 	Phenomenology ---------------------------------------%
%----------------------------------------------------------------------%

\section{NLHT phenomenology}\label{Pheno}

%----------------------------------------------------------------------%
%-------------- Parameter space ---------------------------------------%
%----------------------------------------------------------------------%

\subsection{Parameter space}

\begin{figure}
\centering
\includegraphics[scale=0.6]{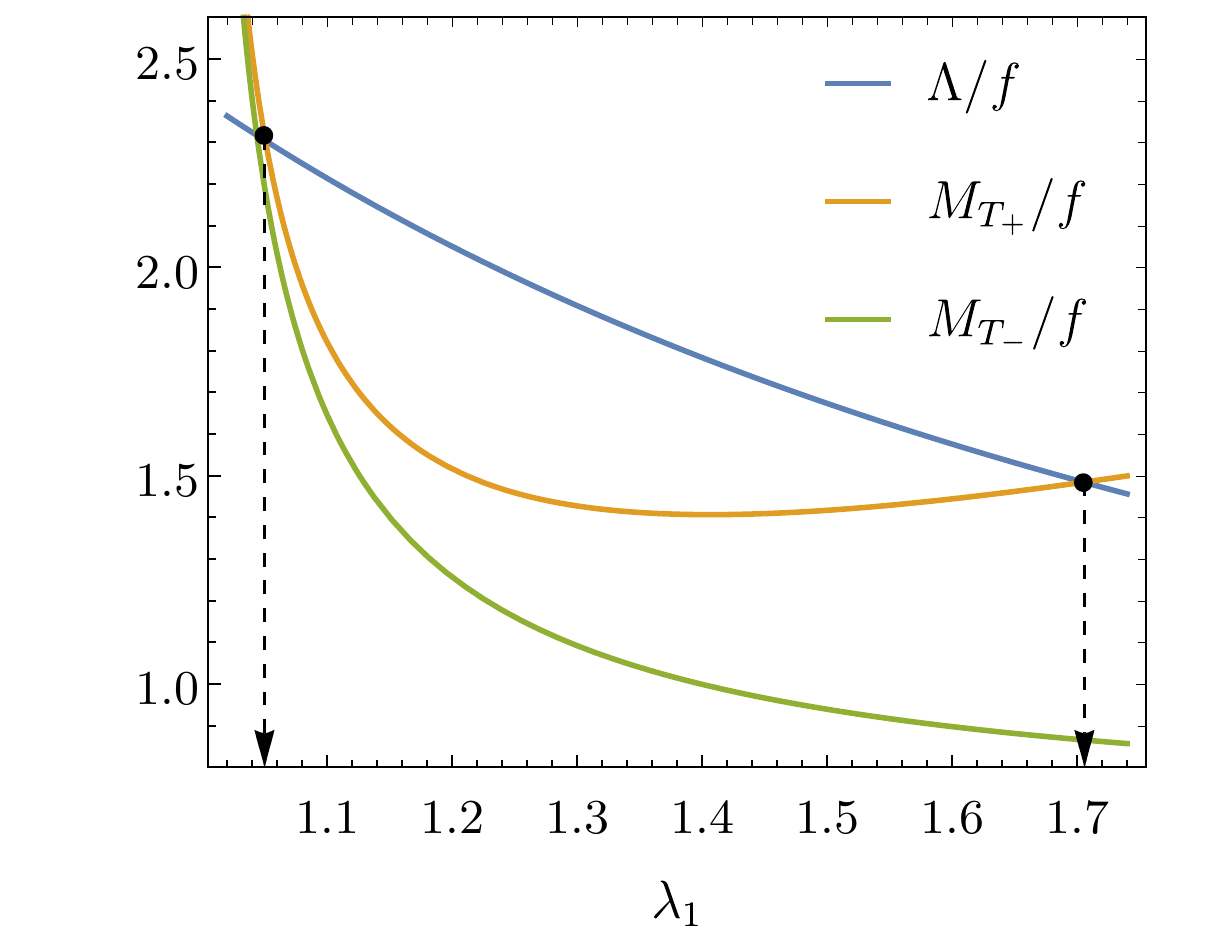} 
\includegraphics[scale=0.6]{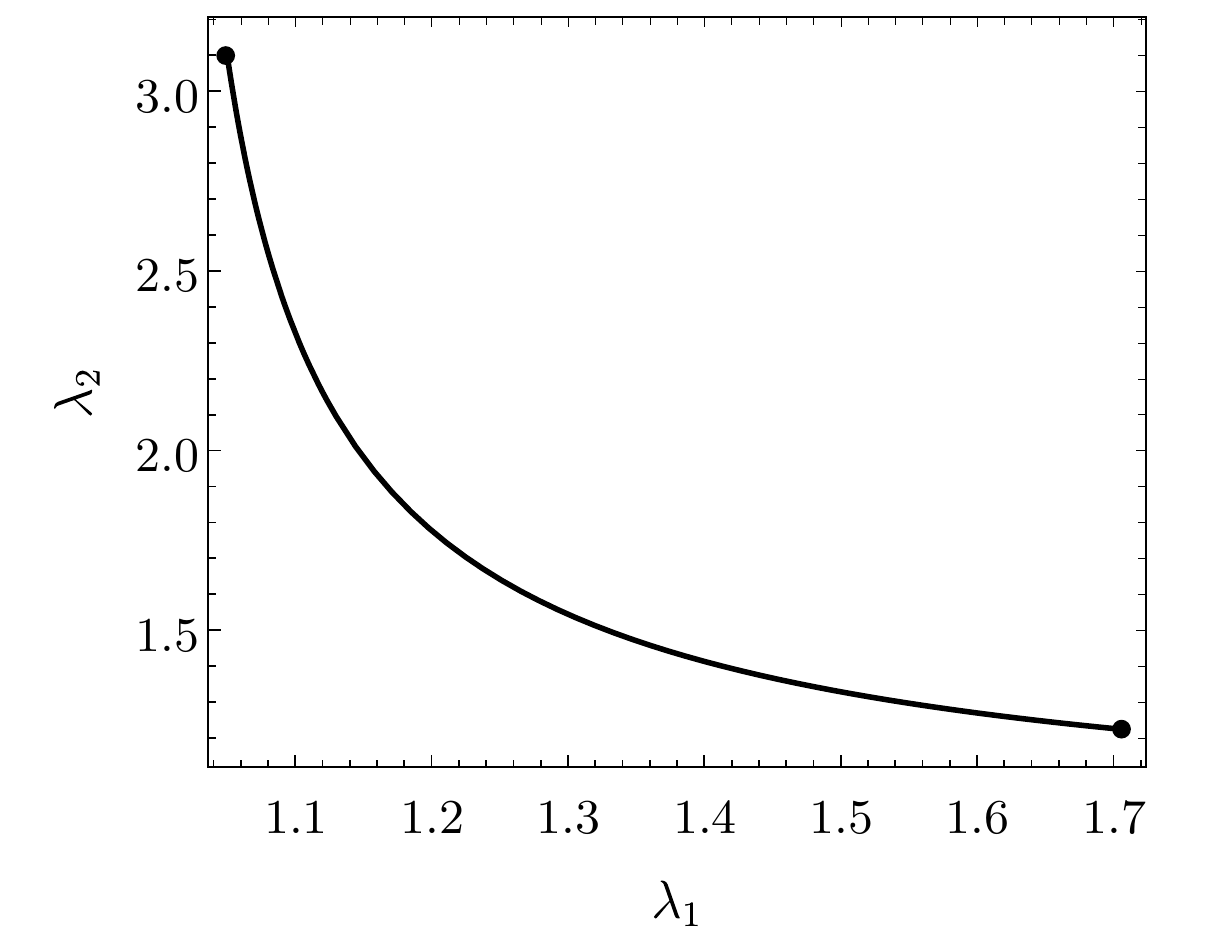} 
\caption{The interval of $\lambda_1$ yielding a top quark partner mass $M_{T_+}$ below the scale $\Lambda$ on the plot of the left-hand side determines the range of possible values of $\lambda_2$ from eq.~(\ref{topmass}) on the plot on the right-hand side. These intervals are independent of $f$ but the mass scales are proportional to $f$.} 
\label{toppartnermass}
\end{figure}

Before constraining the parameters of the model, some relations must be taken into account. Let us focus first on the Yukawa couplings $\lambda_1$ and $\lambda_2$ in eq.~(\ref{toplag}). They are not independent but connected to the top quark mass $m_t\approx173$~GeV \cite{Zyla:2020zbs} according to eq.~(\ref{topmass}). At leading order in $v$ one finds
\eq{
\frac{1}{\lambda_1^2}+\frac{1}{\lambda_2^2}=\left(\frac{v}{\sqrt{2} m_t}\right)^2.
}
The masses of the T-even and T-odd top partners ($T_{+}$ and $T_{-}$, respectively) are given in eq.~(\ref{toppartnermasses}) in terms of these Yukawa couplings. $T_{+}$ is heavier and we have to impose that its mass remains below the cutoff scale $\Lambda$ or otherwise it should be integrated out and would not be part of the spectrum of the theory. The cutoff scale is related to the Higgs quartic coupling $\lambda=\frac{1}{2}(M_h/v)^2\approx 0.13$, the Yukawa coupling $\lambda_1$ and the gauge couplings $g\approx 0.641$ and $g'\approx 0.344$ through eq.~(\ref{higgsquartic}), where $g=e/s_W$ and $g'=e/c_W$ with $s_W^2=1-M_W^2/M_Z^2$ and $e^2=4\pi\alpha$. This constrains the value of $\lambda_1$ to be in the interval $\lambda_1\in \left[1.05, 1.71\right]$, equivalently $\lambda_2\in \left[1.22, 3.10\right]$, as shown in fig.~\ref{toppartnermass}. 
Since the cutoff scale and the top quark partner masses are proportional to $f$, fig.~\ref{toppartnermass} also shows that $\Lambda/f \in \left[1.49, 2.32 \right]$. These values are compatible with the perturbative limit $\Lambda\lesssim 4\pi f$. In addition, current EWPD impose that vector-like quarks must be heavier than about 2~TeV \cite{Vatsyayan:2020jan}. Therefore, we will force the lighter top partner $T_{-}$ to have a mass $M_{T_{-}}>2$~TeV, implying $f\gtrsim900$ GeV.

On the other hand, T-parity is exact and thus the lightest T-odd particle (LTP) is stable. To be a viable dark matter candidate, the LTP must be electrically neutral. In our model, the potentially light and neutral light T-odd particles are the $A_H$ gauge boson, the neutral leptons, with masses proportional to the free parameters $\kappa_l$ and $\widehat{\kappa}_l$, the T-odd singlet $\widehat{\eta}$ and the neutral component of the T-odd real triplet $\widehat{\omega}^0$. The neutral components of the original complex triplet $\Phi$ do not play a role here since they are always heavier than $A_H$ because comparing eqs.~(\ref{gaugebosonmasses}) and (\ref{mphi}) we have that $M_{\phi}/f=\sqrt{2}M_h/v\approx 0.718$ and $M_{A_H}/f=g'\sqrt{2/5}\approx 0.217$. At leading order the neutral leptons share mass with the charged ones and the same happens with the neutral and the charged components of the real triplet. This rules them out. Therefore, the possible dark matter candidates are the singlet $\widehat{\eta}$ and the $A_H$ gauge boson. 

\begin{figure}
\centering
\includegraphics[scale=0.6]{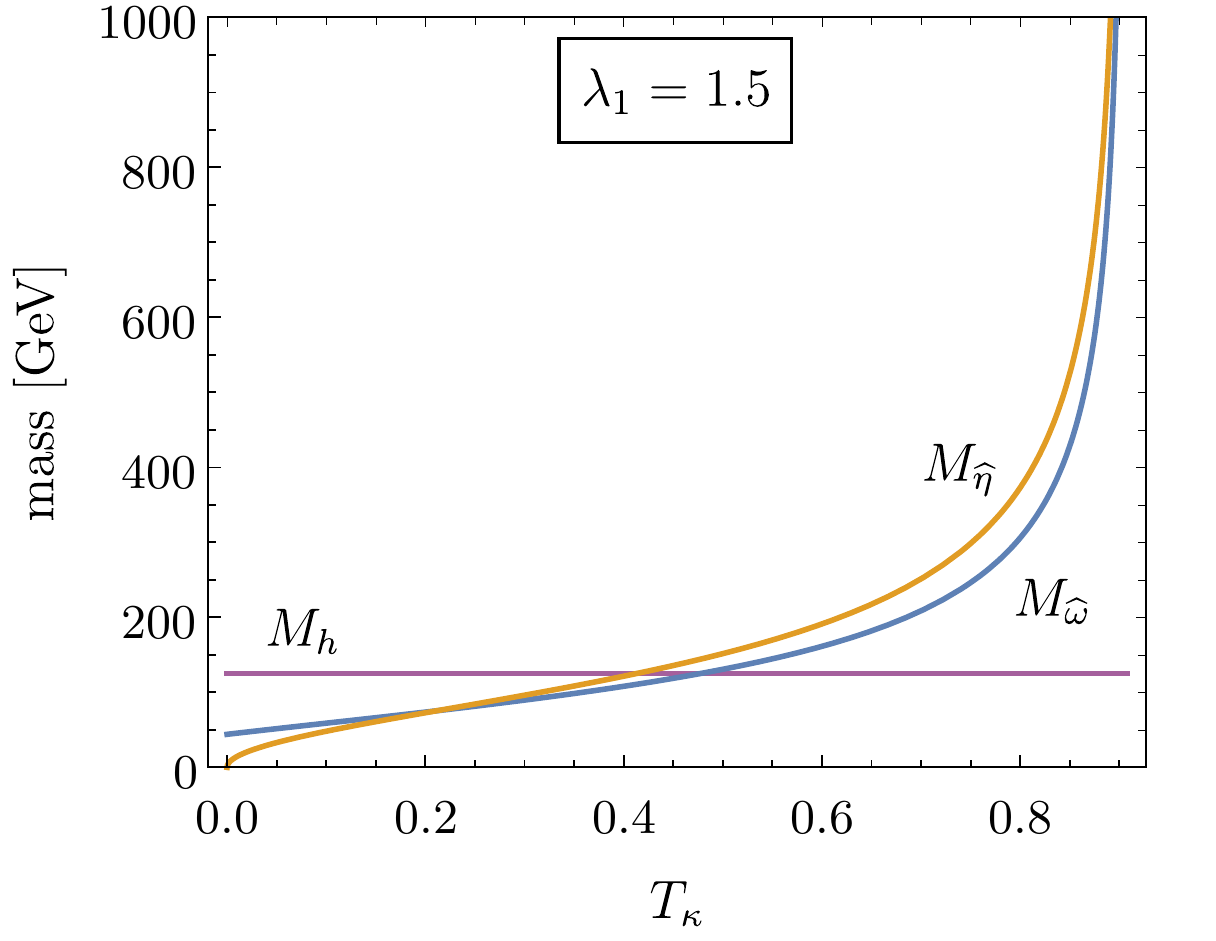} 
\includegraphics[scale=0.6]{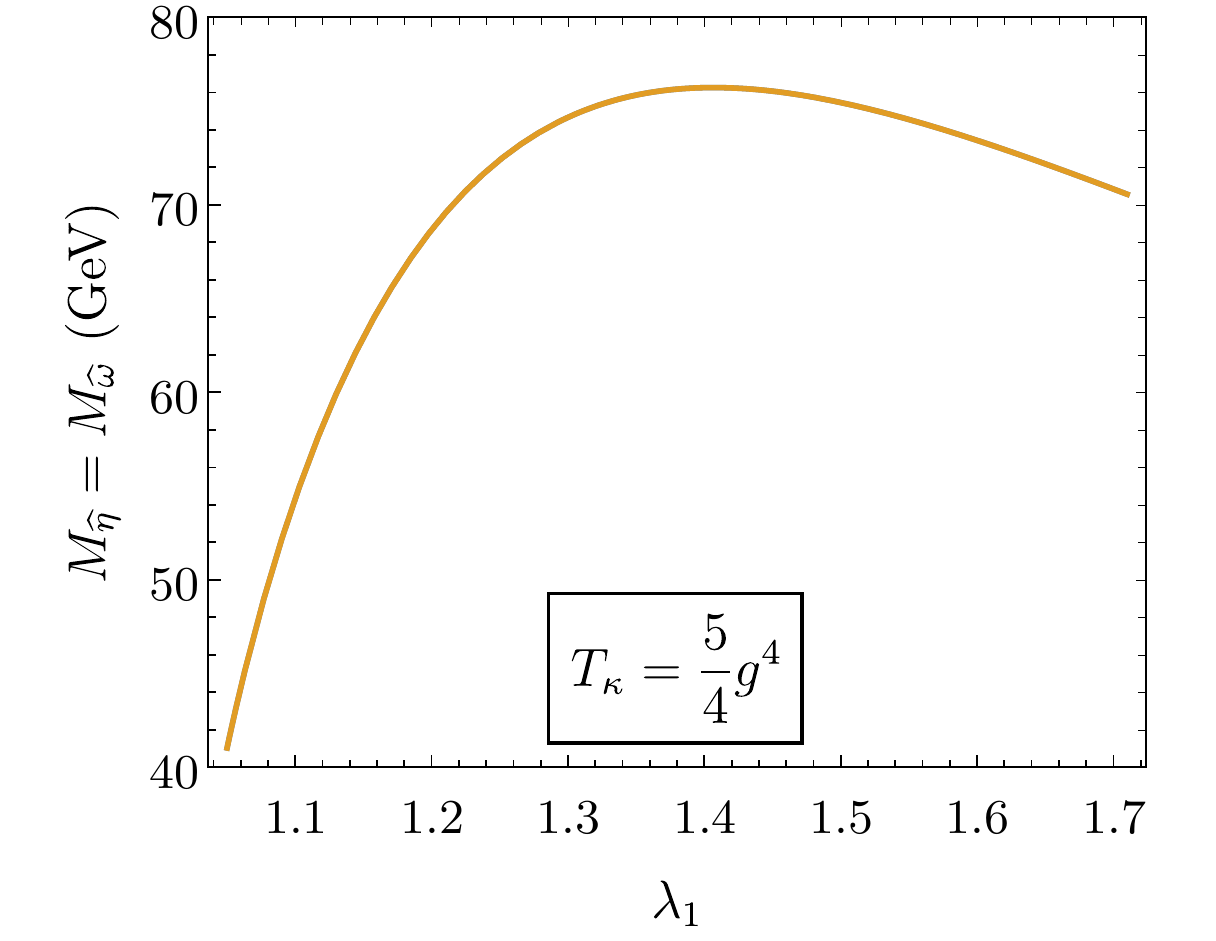} 
\caption{Left: Masses of the Higgs and the new T-odd scalars as a function of $T_{\kappa}$ for a chosen value of $\lambda_1=1.5$. They are naturally light except for $T_{\kappa}$ close to its maximum ($T^{\textrm{max}}_\kappa(\lambda_1=1.5)\approx 0.90$). Right: mass of the T-odd singlet $\widehat{\eta}$ as a function of the allowed values of $\lambda_1$ when $M_{\widehat{\eta}}=M_{\widehat{\omega}}$.} 
\label{metamomega}
\end{figure}

From eq.~\eqref{metaproptomomega} we know that the singlet is lighter than the triplet if $T_{\kappa}<\frac{5}{4}g^4\approx0.211$. Then according to fig.~\ref{metamomega} it has a mass $M_{\widehat{\eta}}\lesssim80$~GeV, and since $M_{A_H}\gtrsim200$~GeV for $f\gtrsim900$ GeV, the singlet would be the LTP if T-odd leptons are heavy enough. This opens a possible window to light scalar dark matter, as already suggested in ref.~\cite{Feng:2014vea}. However, to compare with previous works \cite{hubiszPhenomenologyLittlestHiggs2005,2006,Wang:2013yba}, we will consider the scenario where the heavy photon $A_H$ is the dark matter candidate. Then the new scalars must be heavier than $A_H$, which also implies that the singlet is heavier than the triplet.
 
%----------------------------------------------------------------------%
%-------------- Simplified model --------------------------------------%
%----------------------------------------------------------------------%

\subsection{Simplified model}

It is well known that, in order to reproduce within the LHT the current dark matter relic density $\Omega h^2=0.120\pm 0.001$ \cite{Planck:2018vyg}, it is necessary the presence of co-annihilators if the LTP has $M_{A_H}\gtrsim200$~GeV. In ref.~\cite{2006}, for instance, the T-odd leptons and quarks share masses and are taken nearly mass-degenerate with the $A_H$ gauge boson. However, according to the current constraints on the vector-like quark masses already mentioned, this assumption would lead to a very heavy LTP. Here we will analyze a simplified NLHT model in which all heavy lepton and quark Yukawa couplings are diagonal and degenerate in flavor space, that is, $\kappa_{l,i}=\widehat{\kappa}_{l,i}\equiv \kappa_l$ and $\kappa_{q,i}=\widehat{\kappa}_{q,i}\equiv\kappa_q$, but they have different masses $m_{\ell_H}=\sqrt{2}\kappa_l f\gtrsim M_{A_H}$ and $ m_{q_H}=\sqrt{2}\kappa_q f$.\footnote{Another difference is that the NLHT includes more fermionic degrees of freedom to restore the gauge invariance of the model.} The chosen degeneracy implies that the traces appearing in the expression for the masses of the new T-odd scalars in eqs.~(\ref{mw}) and (\ref{meta}) reduce to $T_{\kappa}=3\kappa_l^4 + 9\kappa_q^4$. We will consider that only the T-odd heavy leptons can act as co-annihilators.

One should keep in mind that the contribution of a co-annihilator to the dark matter relic density is exponentially suppressed as $e^{-\Delta/T_{\rm fo}}$, where $\Delta$ is the mass splitting between the dark matter candidate and the co-annihilator and $T_{\rm fo}\sim 20-30$ GeV is the freeze-out temperature \cite{Griest:1990kh}. To prevent that the scalars influence the relic density, we safely impose a lower bound to the mass of the lighter T-odd real triplet $M_{\widehat{\omega}}> M_{A_H}+ M_h$ using the fact that the new scalar masses can take any value according to fig.~\ref{metamomega}. This defines a minimum for the traces, $T^{\textrm{min}}_{\kappa,\widehat{\omega}}\left(\lambda_1,f\right)$.

\begin{figure}
  \centering
  \includegraphics[width=0.49\linewidth]{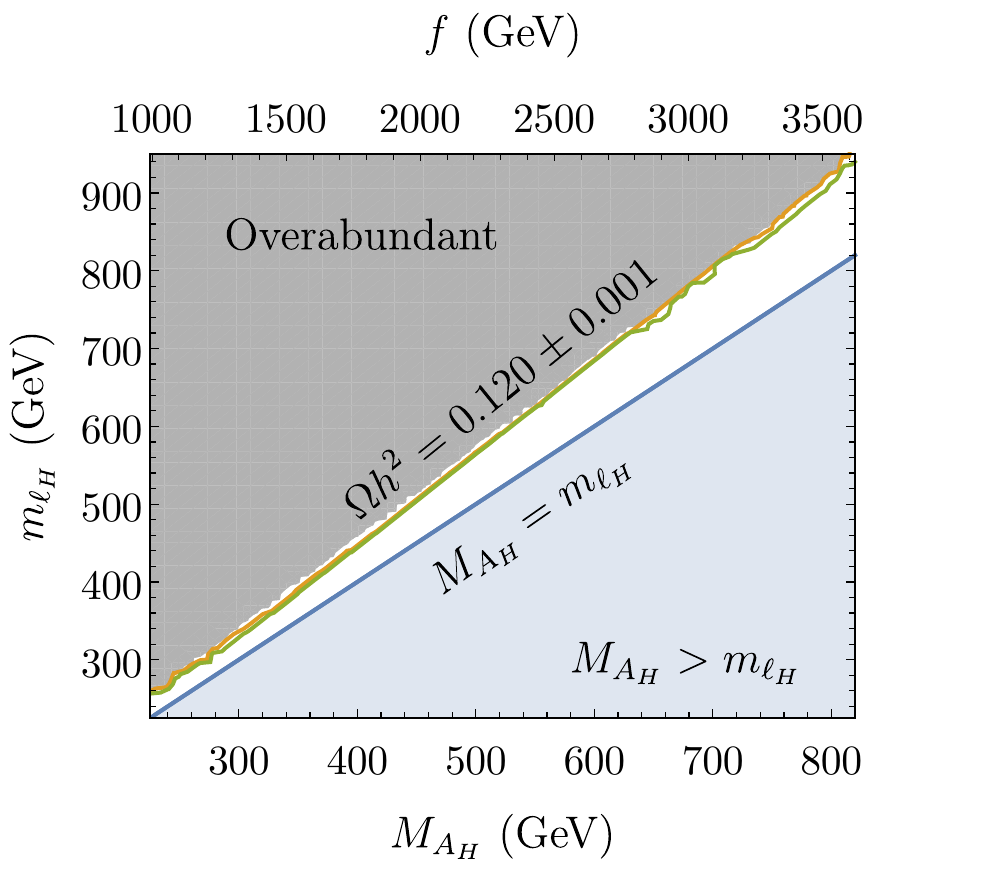}
  \caption{Region in the $M_{A_H}-m_{\ell_H}$ plane compatible with $A_H$ as dark matter candidate (white). The $A_H$ constitutes 100\% (50\%) of the dark matter abundance for masses within the orange (green) bands. The grey region is excluded because it would yield dark matter overabundance and the blue region is excluded because $A_H$ would not be the LTP.} 
  \label{relicdensity}
\end{figure}
  
Implementing our model in FeynRules \cite{Alloul:2013bka}, using micrOMEGAs \cite{Belanger:2013oya} to evaluate the relic density and running T3PS \cite{Maurer:2015gva} to scan the parameter space we obtain the plot in fig.~\ref{relicdensity}. The main contributions to the dark matter relic density come from the co-annihilation of T-odd leptons to pair-produce SM gauge bosons $W^{\pm}, Z$. The narrow orange band covers the region where the relic abundance of $A_H$ is compatible with 100\% of the observed dark matter density, which can be approximated by
\eq{\label{reliceq}
m_{\ell_H} \approx 1.16\; M_{A_H} %-3.60 \mbox{ (GeV)},
}
for $M_{A_H}\in\left[200,800\right]$ GeV. From the expressions of $M_{A_H}$ in eq.~\eqref{gaugebosonmasses} and $m_{\ell_H}=\sqrt{2}\kappa_l f$ the heavy lepton Yukawa coupling is then constrained to $\kappa_l\approx 0.185$ for $f \gtrsim 900$ GeV.

\begin{figure}
  \centering
  \includegraphics[scale=0.6]{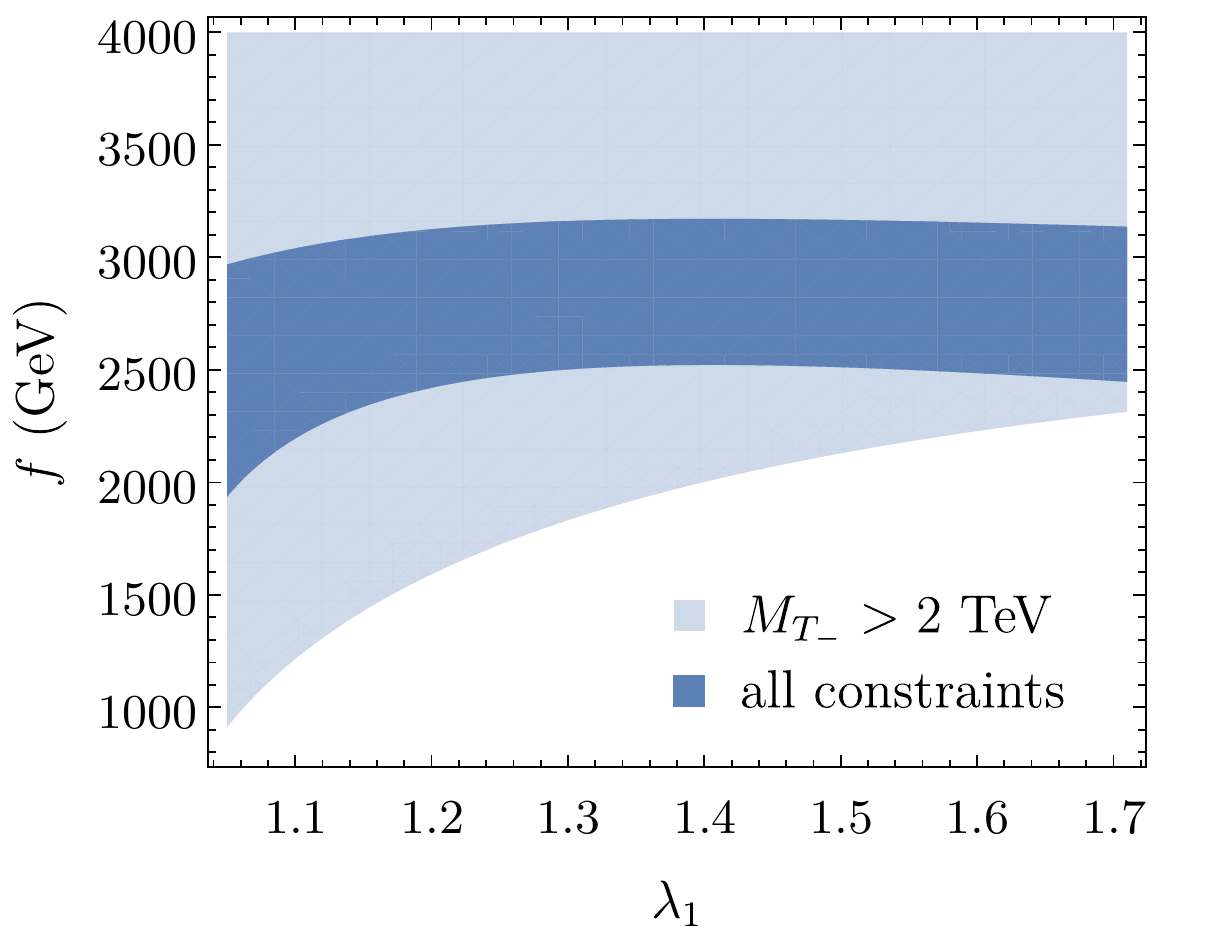} 
  \caption{Values of $f$ compatible with $M_{T_-}>2$~TeV as a function of $\lambda_1$ (light blue). The region is further constrained (darker blue) by the conditions $T^{\textrm{max}}_{\kappa,\widehat{\eta}}\left(\lambda_1\right)>T^{\textrm{min}}_{\kappa,\widehat{\omega}}\left(\lambda_1,f\right)$ (upper bound) and $T_{\kappa,\widehat\eta}^{\rm max}(\lambda_1)> T_{\kappa,q_H}^{\rm min}(f)$ (lower bound) as explained in the text.
  }
  \label{regionfl}
  \end{figure}

We have already set a lower bound to the lighter T-odd scalar mass, $M_{\widehat{\omega}}>M_{A_H}+M_h$. Using naturalness arguments 
one can also expect upper bounds for the scalars. As previously emphasized, the new scalar masses are proportional to the Higgs mass and are independent of the scale $f$, so they should remain light. To be definite we impose $M_{\widehat{\eta}}<1$~TeV, which provides the maximum trace values $T_{\kappa,\widehat\eta}^{\rm max}(\lambda_1)$ for which $M_{\widehat{\eta}}=1$~TeV given $\lambda_1$. Thus the condition $T^{\textrm{max}}_{\kappa,\widehat{\eta}}\left(\lambda_1\right)>T^{\textrm{min}}_{\kappa,\widehat{\omega}}\left(\lambda_1,f\right)$ defines an upper bound on the allowed values of $f$ as a function of $\lambda_1$, which is shown in fig.~\ref{regionfl}. The lower bound that follows from $M_{T_-}>2$~TeV is also depicted.
  
The condition that $A_H$ is the LTP implies $m_{q_H}/f> M_{A_H}/f$, that is $\kappa_q>g'/\sqrt{5}\approx 0.16$. However the current bound on the heavy quark masses sets a limit $\kappa_q > \sqrt{2}\mbox{ TeV}/f$, which is stronger for the values of $f\lesssim3$~TeV allowed by fig.~\ref{regionfl}. From this and the already constrained value of $\kappa_l\approx 0.185$ we derive the restriction $T_{\kappa}>T_{\kappa,q_H}^{\rm min}(f)\equiv 0.0035+36\;\mbox{TeV}^4/f^4$. This additional condition for the traces finally sets the allowed region in the $\lambda_1-f$ plane of fig.~\ref{regionfl}, which features a possible window for $f$ between 2.0 and 3.1~TeV. The minimum for the traces is actually given by $T^{\textrm{min}}_{\kappa,\widehat{\omega}}\left(\lambda_1,f\right)$.

%----------------------------------------------------------------------%
%-------------- Spectrum and scalar decays  ---------------------------%
%----------------------------------------------------------------------%

\subsection{Particle spectrum and scalar decays} 

\begin{figure}
\centering
\includegraphics[scale=.6]{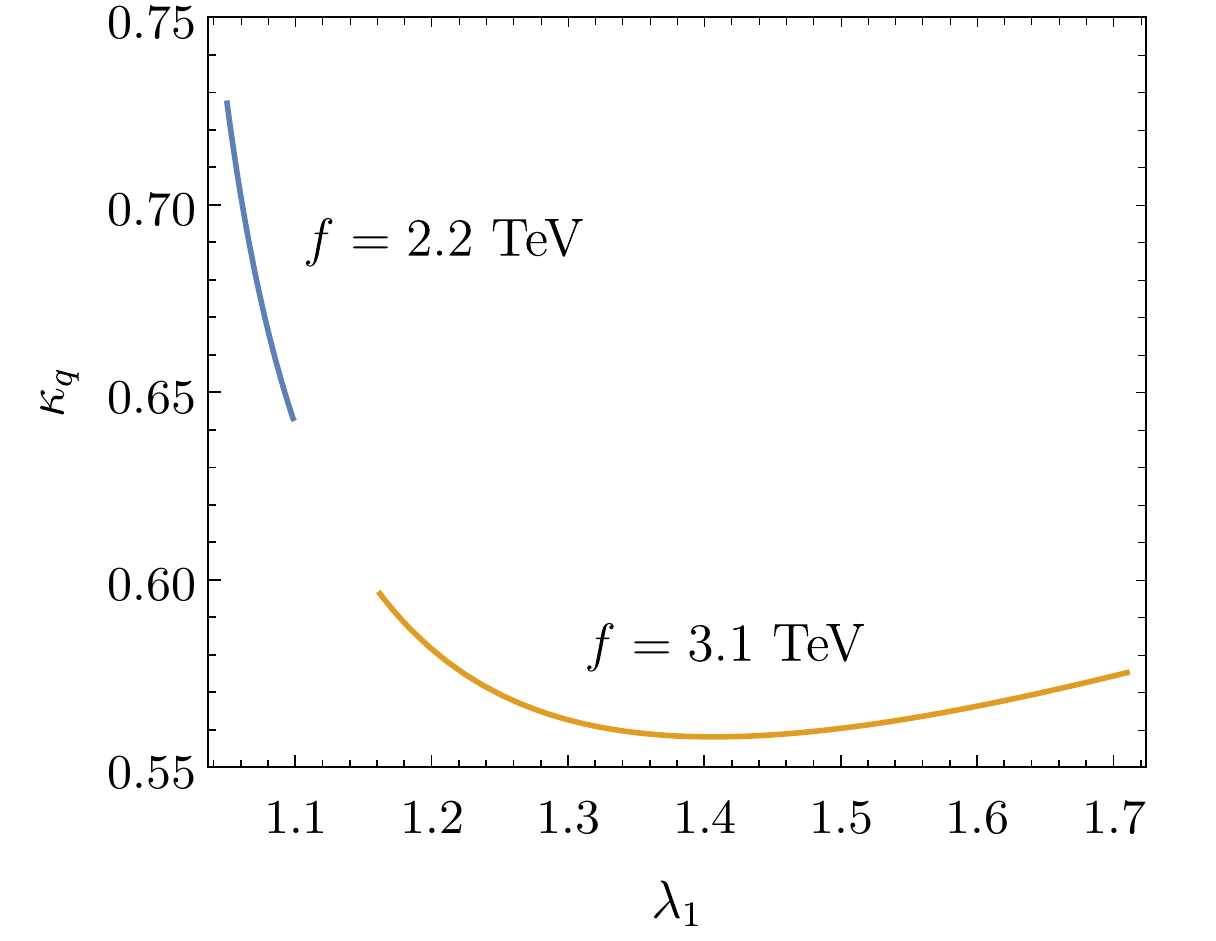}
\includegraphics[scale=.6]{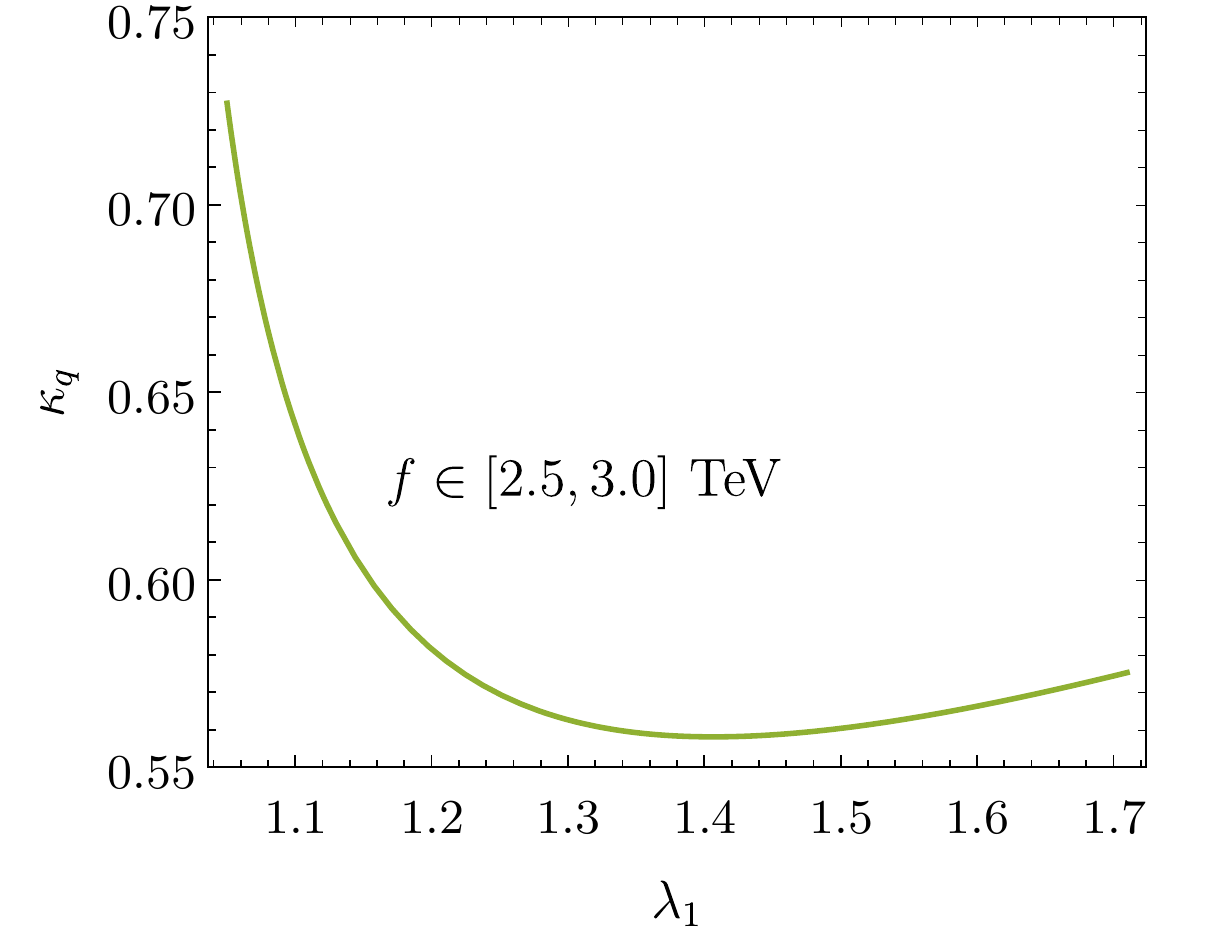}
\caption{Correlation between the heavy quark Yukawa coupling $\kappa_q$ and the top quark Yukawa coupling $\lambda_1$ for different values of the scale $f$.} 
\label{correlation}
\end{figure}

As a consequence of all previous constraints, the Yukawa coupling of the heavy quarks $\kappa_q$ and $\lambda_1$ get strongly correlated (fig.~\ref{correlation}). This correlation is nearly independent of $f$. This is because the maximum and minimum values of $\kappa_q$ are extremely close due to the asymptotic behavior of the new scalar masses with $T_{\kappa}$ and $T_{\kappa,\widehat{\eta}}^{\max}$ does not depend on $f$. However, not all values of $\lambda_1$ are available for every possible value of $f$, except for $f\in[2.5,3.0]$, as can be seen in fig.~\ref{regionfl} and is reflected on fig.~\ref{correlation}. Comparing figs.~\ref{toppartnermass} and \ref{correlation} one can explicitly check that heavy quarks are safely below the cut-off scale because $m_{q_H}/f=\sqrt{2}\kappa_q<\Lambda/f$. Their common masses are in practice a function of $\lambda_1$, ranging between $0.8 f$ and $f$. 

\begin{figure}
  \centering
 \includegraphics[scale=.6]{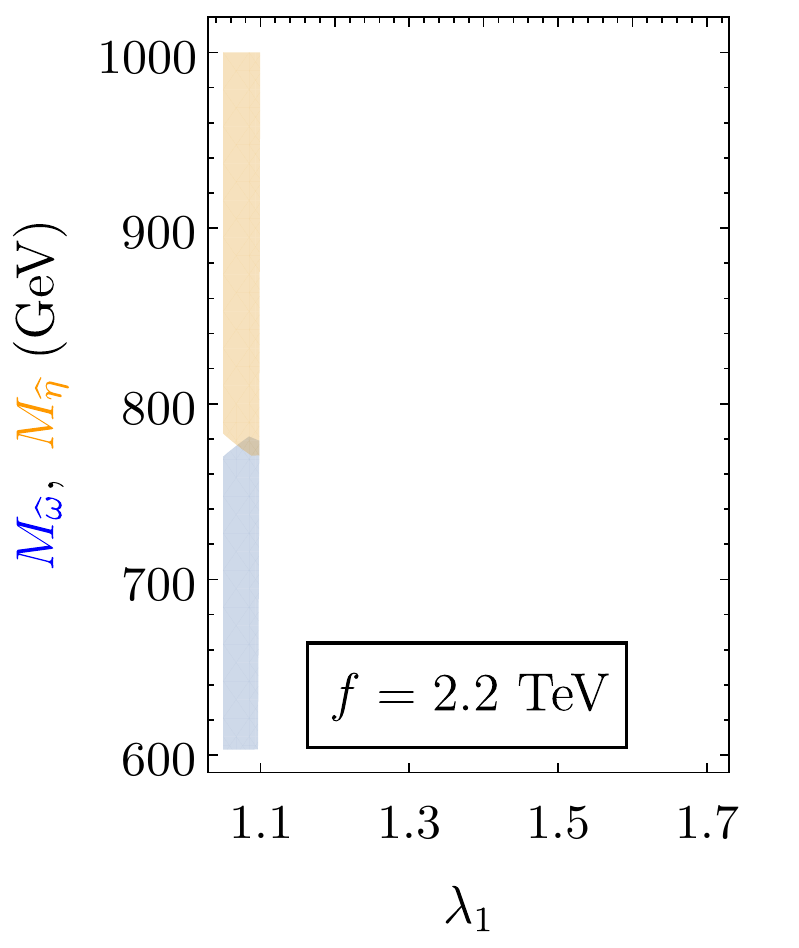}
  \includegraphics[scale=.6]{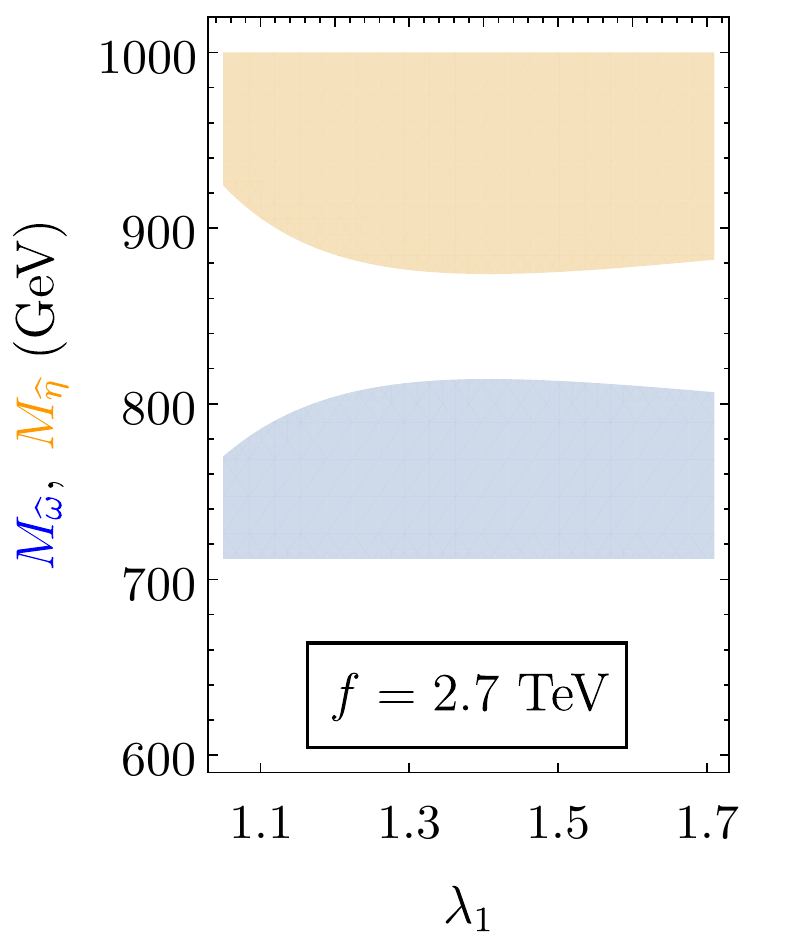}
  \includegraphics[scale=.6]{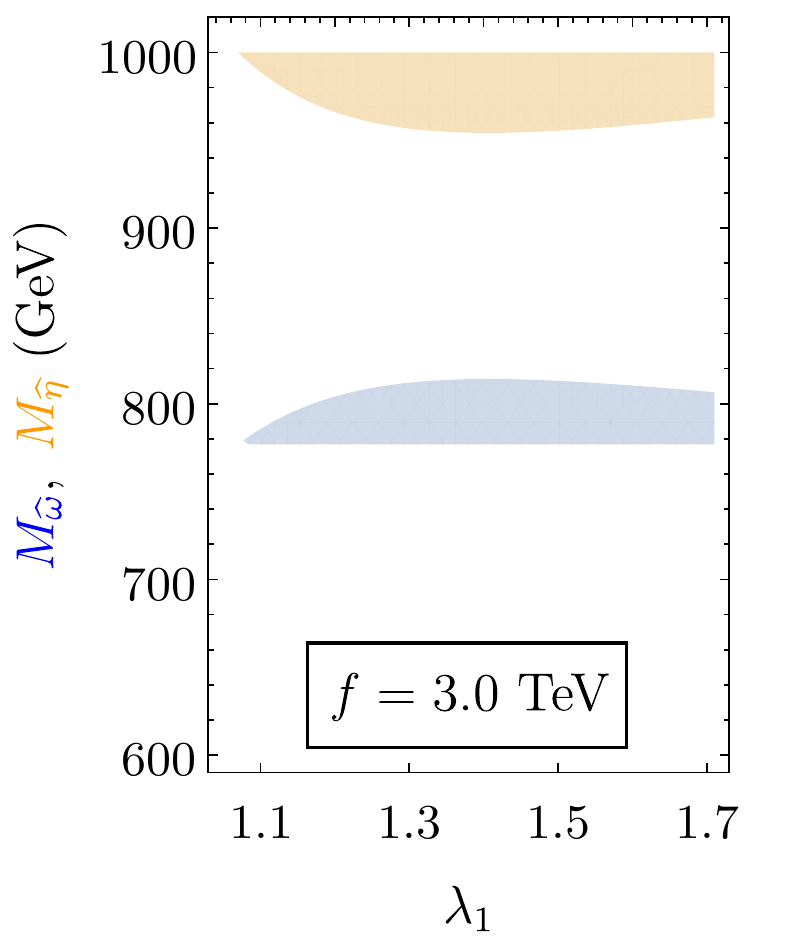}
  \caption{Ranges of allowed masses of the new T-odd scalars for different values of the scale $f$ as a function of $\lambda_1$} 
  \label{scalarmasses}
\end{figure}

\begin{figure}
  \centering
  \includegraphics[width=0.55\linewidth]{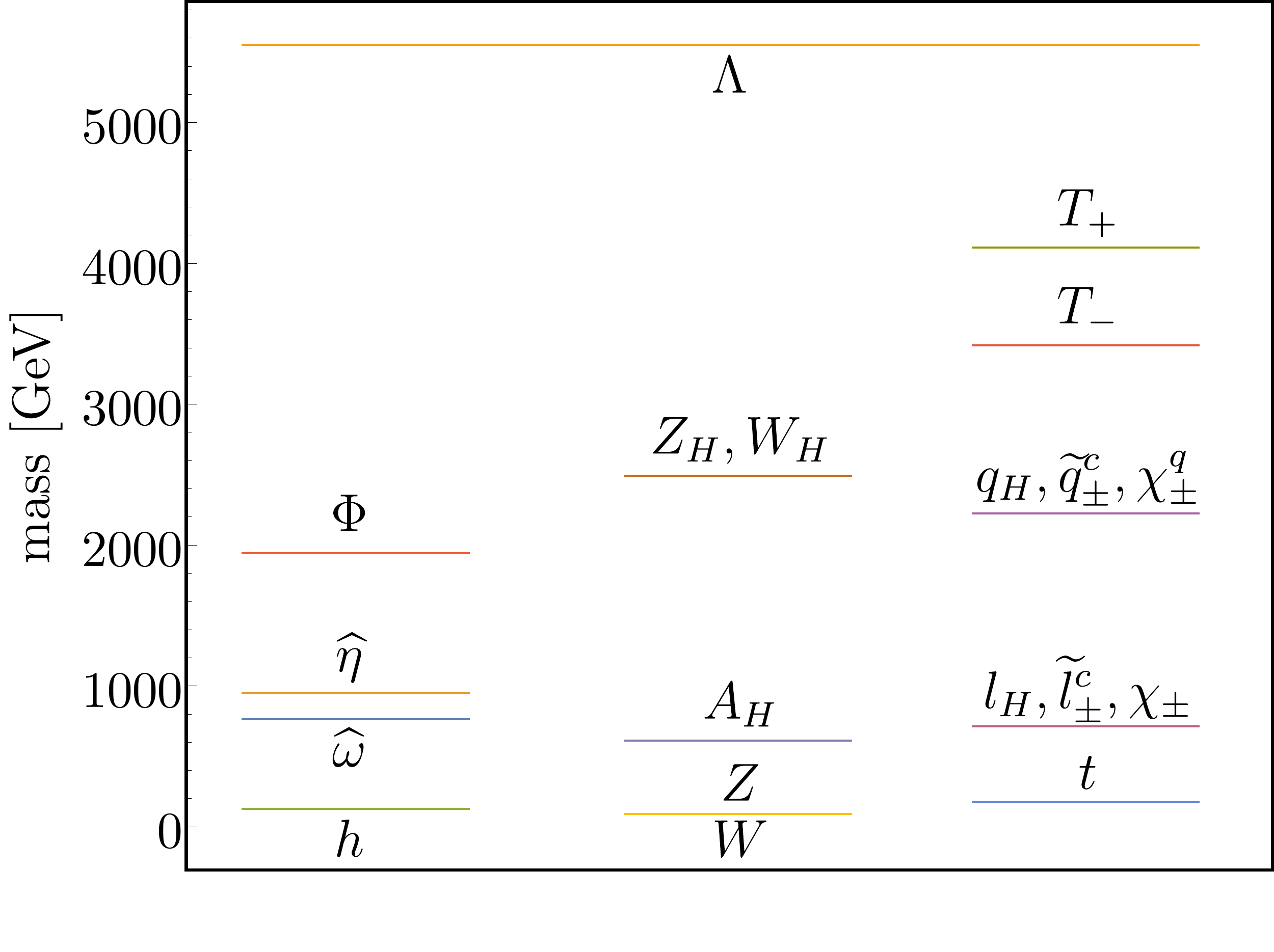}
  \caption{Typical NLHT spectrum of our simplified model with $f\approx 2.7$ TeV and $\lambda_1=1.2$.} 
  \label{spectrum}
\end{figure}

Regarding the scalar masses, fig.~\ref{scalarmasses} shows that they are constrained to the intervals $M_{\widehat{\omega}}\in \left[600, 800\right]$~GeV and $M_{\widehat{\eta}}\in \left[800, 1000\right]$~GeV. The maximum values are fixed by $M_{\widehat{\eta}}<1$~TeV. The minimum values come from the condition $M_{\widehat{\omega}}> M_{A_H}+M_h$ depending on $f$. The allowed values for $\lambda_1$ are determined by $T_{\kappa,\widehat\eta}^{\rm max}(\lambda_1)> T_{\kappa,q_H}^{\rm min}(f)$ and $T_{\kappa,\widehat\eta}^{\rm max}(\lambda_1)>T_{\kappa,\widehat\omega}^{\rm min}(\lambda_1,f)$. For completeness we show below the mass ranges of the T-odd gauge bosons, the usual T-odd complex scalar triplet and the heavy leptons for the allowed values of the scale $f$,
\begin{align}\label{fieldsmass}
  M_{A_{H}}&\in \left[450,680\right] \textrm{ GeV},\\
  M_{W_{H}}=M_{Z_{H}}&\in \left[1850, 2750\right] \textrm{ GeV},\\
  M_{\Phi}&\in \left[1450,2150\right] \textrm{ GeV},\\
  m_{\ell_H}&\in \left[530, 800\right] \textrm{ GeV}.
\end{align}

Our vector-like leptons decay to one standard lepton and a heavy photon, an exotic decay that is not excluded in principle by current LHC searches within these mass ranges \cite{Guedes:2021oqx}.
For $f\approx 2.7$ TeV, the typical spectrum is shown in fig.~\ref{spectrum}. The top quark partners are always the heaviest particles.  The rest of heavy quarks (T-even and T-odd quarks share masses) are lighter but above 2 TeV to comply with EWPD constraints. Apart from the top quark partners, the T-odd $W_H$ and $Z_H$ are the heaviest gauge bosons with a large gap with the $A_H$ gauge boson, the LTP. The heavy leptons are only a few tens of GeV heavier than $A_H$. This is because, as in the quark case, these T-even and T-odd leptons are mass degenerate in our simplified model and the latter act as co-annihilators to reproduce the dark matter relic density. Finally, in the scalar sector, the usual T-odd triplet $\Phi$ is always the heaviest one, followed by the singlet $\widehat{\eta}$ and the triplet $\widehat{\omega}$. 
  
Now that we have studied the full model spectrum, let us discuss the decay channels of the new T-odd scalars $\widehat{\omega}$ and $\widehat{\eta}$. At leading order they decay mostly into a left-handed SM lepton and a right-handed T-odd mirror lepton preserving both T-parity and electric charge. The couplings involved are independent of the Higgs \vev\ because the SM left-handed leptons are SU(2) doublets and so are their mirror versions with same hypercharge as the SM leptons. They are proportional to $\kappa_\ell$ and scale as
\eq{
  \widehat\omega^\pm:\widehat\omega^0:\widehat\eta \sim 1/\sqrt{2}:1/2:1/\sqrt{20}.
}
Thus, the quantum numbers can combine to give those of a triplet for $\widehat{\omega}$ and a singlet for $\widehat{\eta}$. The decays to other leptons with opposite T-parities involve couplings suppressed by powers of $v/f<0.12$ or are kinematically suppresed by the higher masses, of at least $m_{\ell_H}\approx 530$ GeV in our simplified model. In addition, from eq.~\eqref{lagS}, the charged components of the triplet $\widehat\omega^\pm$ can decay into the $A_{H}$ gauge boson and a SM $W^{\pm}$ while the neutral component $\widehat\omega^0$ can decay into a $A_{H}$ and a Higgs boson. The singlet $\widehat\eta$ can decay in the same way as the neutral component of the triplet but, being heavier, it can also decay to the neutral component of the triplet and a Higgs boson. However those couplings are also suppressed by powers of the Higgs \vev. We have explicitly checked that these other channels contribute less than a 3\% to the total decay widths. 

\begin{figure}
  \centering
  \includegraphics[width=0.49\linewidth]{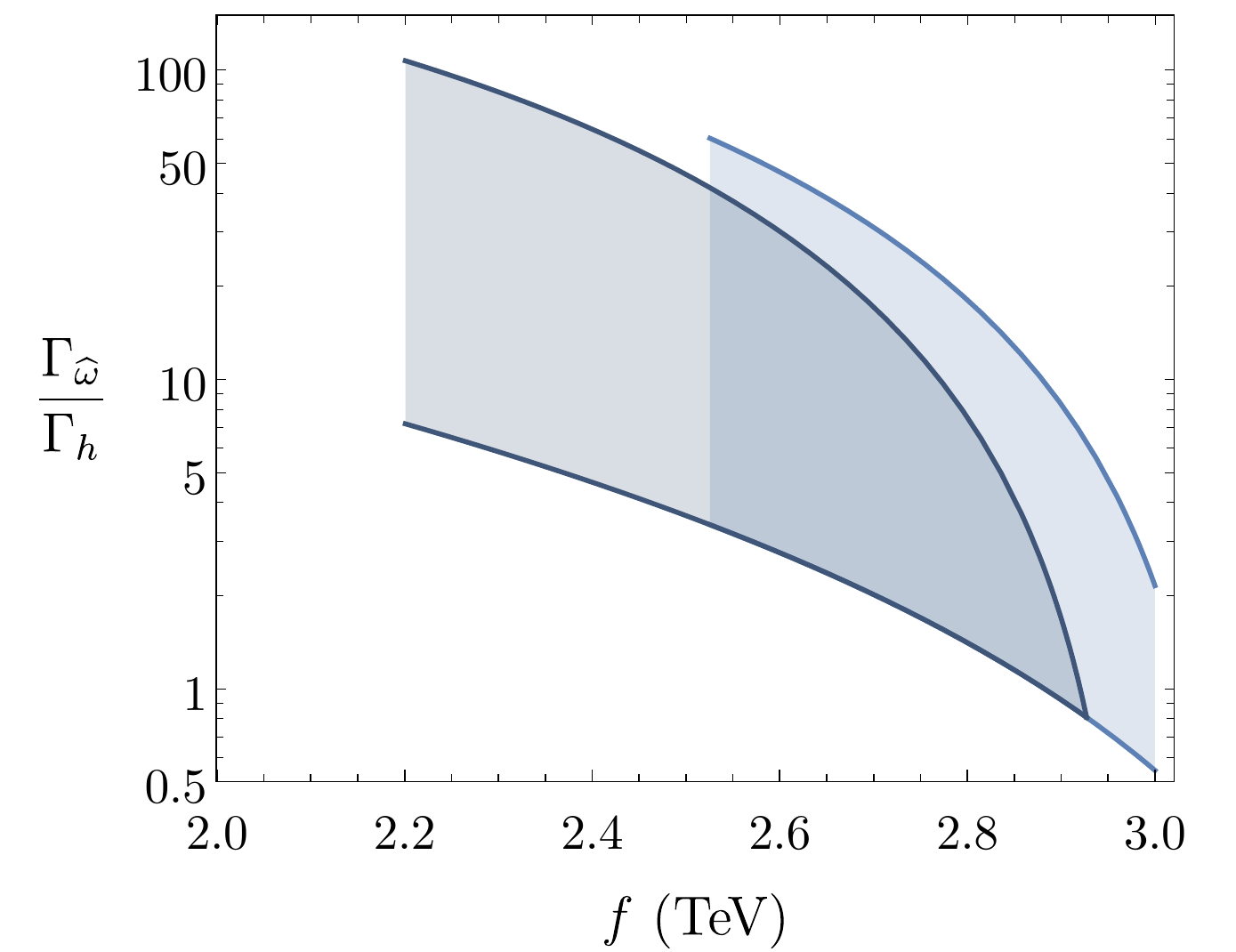}
  \includegraphics[width=0.49\linewidth]{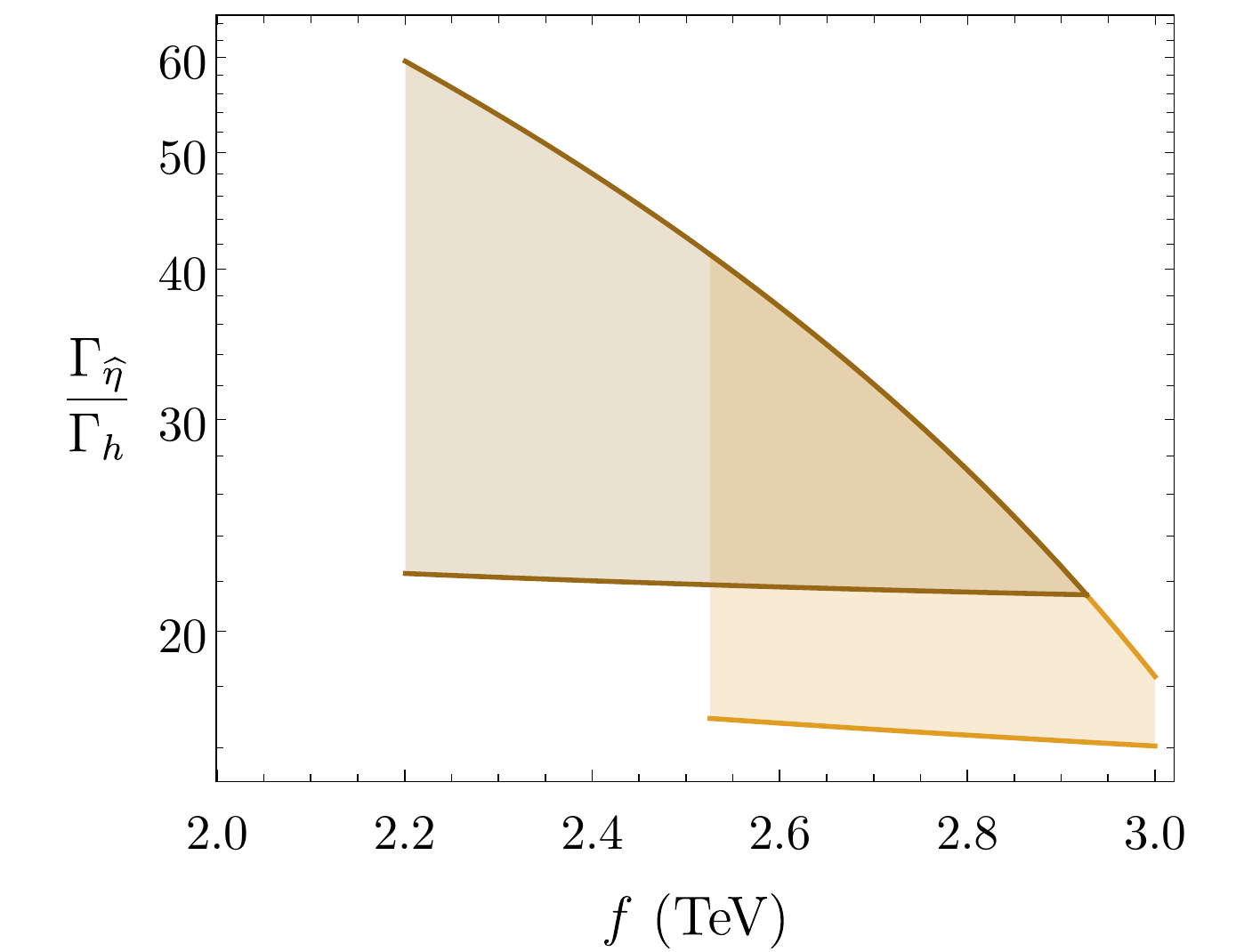}\hfill
  \caption{Decay widths for the triplet and singlet scalar fields normalized to the Higgs width for $\lambda_1=1.1$ and $\lambda_1=1.5$ (lighter color). The upper and lower bounds correspond to the maximum and minimum values of the scalar masses.}
  \label{decaywidths}
\end{figure}
  
In fig.~\ref{decaywidths} we show the range of values of the triplet and singlet decay widths normalized to the Higgs boson width for a couple of values of $\lambda_1$. The charged and neutral components of the triplet have the same width at this order because the couplings to the neutral component are a factor $1/\sqrt{2}$ smaller than those for the charged components but for the neutral component one can exchange particles for antiparticles in the final state adding a contribution that compensates the factor $(1/\sqrt{2})^2$ in the decay width. Note that the upper bounds for the width of triplet and singlet are of the same order. This is because the singlet is heavier than the triplet but the couplings of the singlet to leptons are a factor $\sqrt{5}$ smaller than those for the neutral component of the triplet. On the other hand, the lower bound of the triplet decay width is much smaller. This suppression comes from the kinematic factor since the lower bound for the triplet mass is only a few tens of GeV larger than the mass of the T-odd mirror leptons. In any case their lifetimes are small, since in the worst case the triplet could live twice more than the Higgs in the available parameter space.

Nevertheless, these new scalar particles would not be generated in a significant amount at the LHC. They are produced by an electroweak interaction together with another T-odd particle and the energy threshold for this process is very high considering the usual spectrum in fig.~\ref{spectrum}. 

%-----------------------------------------------------------------------%
%--------------- CONCLUSIONS -------------------------------------------%
%-----------------------------------------------------------------------%

\section{Conclusions}\label{Conclusions}

In this work we have reviewed the new Littlest Higgs model with T-parity (NLHT) \cite{illana2022new} and explored its main phenomenological consequences. This model was developed to cure a non gauge invariance issue in the fermion sector of the original LHT. For this purpose the global symmetry group $\su5$ was extended with an $\left[\su2\times \u1\right]^2$ factor. The global group gets spontaneously broken to $\so5\times\left[\su2\times \u1\right]$ at a high energy scale $f$. The spontaneous breaking of the extra piece of the global group gives rise to four new T-odd scalars. Only a subgroup $\left[\su2\times\u1\right]_1\times \left[\su2\times\u1\right]_2$ of the full global group is gauged, that gets spontaneously broken to the SM gauge group by the same \vev\ as the global group. 

Under the gauge group, the four new T-odd scalars share quantum numbers with some of the would-be Goldstone bosons of the original LHT: a triplet with zero hypercharge and a singlet. As a consequence they mix. After field redefinitions, some combinations become the would-be Goldstone bosons of the NLHT, absorbed by the heavy gauge bosons, and the rest are the usual triplet $\Phi$ and a new triplet $\widehat{\omega}$ and a singlet $\widehat{\eta}$, which are physical. Their corresponding masses are obtained evaluating the gauge boson and the heavy fermion contributions to the Coleman-Weinberg potential. The new scalars have masses proportional to the Higgs mass, independent of the scale $f$. Parametrically, apart from gauge couplings, the new scalar masses depend on the Yukawa couplings $\lambda_1$ and $\lambda_2$, that provide masses to the top quark and its corresponding T-even and T-odd partners, and the Yukawa couplings $\kappa$ and $\widehat\kappa$, giving gauge invariant mass terms to the rest of non standard heavy fermions (in principle different for quarks and leptons). These are more than in the original model and have either of the T-parities.

The top quark Yukawa coupling $\lambda_2$ is a function of $\lambda_1$ given the top quark mass. The condition that the heavier top quark partner is below the cutoff scale $\Lambda$ constrains the value of $\lambda_1$ to the interval $\left[1.05, 1.71\right]$. And the experimental lower bound on vector-like quark masses above 2~TeV \cite{Vatsyayan:2020jan} pushes the allowed value of $f\gtrsim900$~GeV.

Since T-parity is exact, the LTP is stable. To have a viable dark matter candidate, the LTP must be electrically neutral. The NLHT contains two potential candidates, the singlet $\widehat{\eta}$ and the usual heavy photon $A_H$. Although there is enough space to explore the singlet as a dark matter component (with $M_{\widehat\eta}\lesssim80$~GeV), we have chosen it to be the heavy photon in order to compare with previous works \cite{hubiszPhenomenologyLittlestHiggs2005,2006,Wang:2013yba}. The aforementiond lower bound to $f$ implies that the LTP has $M_{A_H}\gtrsim200$~GeV and $M_{\widehat\omega}< M_{\widehat\eta}$. 
On the other hand, the scalar masses do not depend on the high energy scale $f$ and thus they must remain light by naturalness arguments. Therefore, to be conservative, we impose an upper bound of 1 TeV to the mass of the singlet.

In order to reproduce the current relic density with heavy photons, the presence of co-annihilators is necessary unless the LTP is lighter than the lower bound set above \cite{2006}. To put vector-like quarks beyond current bounds and at the same time provide a not so heavy dark matter candidate yielding the right relic abundance, we have adopted a simplified NLHT model with degenerate and flavor diagonal Yukawa couplings $\kappa=\widehat\kappa$, different for leptons ($\kappa_\ell$) and quarks ($\kappa_q$), such that the corresponding heavy masses are $m_{\ell_H}=\sqrt{2}\kappa_\ell f\gtrsim M_{A_H}$ and $m_{q_H}=\sqrt{2}\kappa_q f\gtrsim2$~TeV, respectively. We also require that only T-odd leptons act as co-annihilators by taking $M_{\widehat{\omega}}>M_{A_H}+M_h$.  

As a consequence of all these constraints, the allowed high energy scale $f$ lies approximately in the interval $f\in\left[2,3\right]$ TeV, the common Yukawa coupling of the heavy quarks gets strongly correlated to the top Yukawa coupling $\lambda_1$ and, demanding that all dark matter of the universe is made of the NLHT heavy photons, the Yukawa coupling for all heavy leptons is fixed to $\kappa_l\approx 0.185$. A typical spectrum of this scenario is displayed in fig.~\ref{spectrum}.

Finally we studied the dominant decay channels of the new scalar particles. We found that $\widehat\omega$ and $\widehat\eta$ decay mostly into a T-odd mirror lepton and a SM lepton with the proper quantum numbers. Other channels are negligible because they involve couplings that suffer from suppressions by powers of $v/f<0.12$. They decay very fast; the (lighter) triplet $\widehat\omega$ could live at most two times longer than the Higgs boson. In any case, these new scalars are heavier than about 600~GeV and would be generated by and electroweak interaction together with another heavy T-odd particle at the LHC, so not very sizeable production rates are expected. 

As a future research it would be interesting to further develop the idea of the scalar singlet $\widehat{\eta}$ acting as dark matter candidate. Phenomenologically there is still a window for light scalar dark matter \cite{Feng:2014vea} and the NLHT model has enough room to accommodate this scenario as well.

%-----------------------------------------------------------------------%
%---------------- ACKNOWLEDGMENTS --------------------------------------%
%-----------------------------------------------------------------------%
\section*{Acknowledgments}

We would like to thank M.~Chala, A.~Djouadi, T.~Hahn, M.~Masip and J.~Santiago for very helpful discussions. 
This work was supported in part by the Spanish Ministry of Science, Innovation and Universities (PID2019–107844GB-C21/AEI/10.13039/501100011033), and by Junta de Andalucía (FQM 101, SOMM17/6104/UGR, P18-FR-5057). 

%Funding for open access charge: Universidad de Granada / CBUA.

\bibliography{biblio}

\end{document}